\documentclass[twocolumn,nofootinbib,showpacs,prd]{revtex4}%
\usepackage{amsmath}
\usepackage{amsfonts}
\usepackage{amssymb}
\usepackage{graphicx}

\begin{document}

\title{Non-Equilibrium Dynamics of a Noisy Quantum Ising Chain: statistics of the work and prethermalization after a sudden quench of the transverse field}

\author{
Jamir Marino${}^{1,2}$ and Alessandro Silva${}^{1,3}$}

\address{$^1$ SISSA, – International School for Advanced Studies, via Bonomea 265, 34136 Trieste, Italy}
\address{$^2$ INFN, – Istituto Nazionale di Fisica Nucleare, sezione di Trieste}
\address{${}^3$ Abdus Salam ICTP, Strada Costiera 11, 34151 Trieste, Italy}

\begin{abstract}
We discuss the non-equilibrium dynamics of a Quantum Ising Chain (QIC) following a quantum quench of the transverse field and in the presence of a gaussian time dependent noise.
We discuss the probability distribution of the work done on the system both for static and dynamic noise. While the effect of static noise is to smooth the low energy threshold of the statistic of the work, appearing for sudden quenches, a dynamical noise protocol  affects also the spectral weight of such features. We also provide a detailed derivation of the kinetic equation for the Green's functions on the Keldysh contour and the time evolution of observables of physical interest, extending previously reported results (J. Marino, A. Silva, Phys. Rev. B 86, 060408 (2012)), and discussing the extension of the concept of \emph{prethermalization} which can be used to study noisy quantum many body hamiltonians driven out-of-equilibrium.

\end{abstract}

\pacs{05.70.Ln, 75.10.Jm, 05.30.Jp, 71.10.Fd}

\date{\today}

\maketitle

\section{Introduction}

In the last decade a series of ground-breaking experiments on the dynamics of cold atoms \cite{Greiner} have generated new interest in the thermalization dynamics of quantum many-body systems. If a quantum many-body system is prepared in the ground state of a given hamiltonian $H_i$ and evolved according to a new hamiltonian $H_f$, it is natural to ask whether the excess energy will redistribute among the elementary degrees of freedom and whether the system will eventually reach the thermal state at later times (in the thermodynamic limit). This expectation has been tested theoretically in pioneering works on this subject \cite{Srednicki}, partially confirming the idea that non-integrable quantum many-body systems thermalize, in the sense that observables of physical interest appear to reach asymptotically the value predicted by the Gibbs Ensemble with a temperature set by the energy injected in the system. A noticeable exception are integrable quantum many-body systems, which relax towards a Generalized Gibbs Ensemble (GGE), i.e. a grandcanonical ensemble which takes into account all the conserved quantities of the system \cite{Barthel}.

The simplest protocol to study non-equilibrium dynamics is the so-called \emph{quantum quench}, which consists in preparing the system in the ground state of a quantum many body hamiltonian $H(g_0)$, and let them evolve according to a different hamiltonian $H(g)$, the control parameter $g_0$ being suddenly switched to $g$. Though, the most recent developments in out-of-equilibrium dynamics of quantum many body systems have been mainly concerned in understanding which is the asymptotic steady state attained after a quantum quench, (for a complete review on this subject, see for instance \cite{Polkovnikov2011}), it is still not clear what are the time scales of thermalization, whether the process of thermalization is sudden or composed by many stages, and which are the mechanisms behind thermalization in quantum many-body systems.


Recent theoretical studies of quantum many body systems weakly perturbed away from integrability suggests that first the system relaxes towards a \emph{pre-thermal state}, where the expectation values of observables are predicted by a modified GGE (strongly influenced by the close integrable point ~\cite{Kollar2011}) and only later when inelastic scattering becomes relevant the system departs from the pre-thermal state approaching the asymptotic thermal state. This phenomenon known as \emph{prethermalization} has been studied in many systems of physical interest, ranging from quantum field theories \cite{Berges2004}, to the Hubbard model \cite{Moeckel2010}, Luttinger liquids \cite{Mitra2013}, spinor condensates~\cite{Barnet2011} and non-integrable versions of the Quantum Ising Chain \cite{Marcuzzi}.  Signatures of this crossover have been observed experimentally in split one dimensional condensates~\cite{Gring2011}. While it is evident that the dynamics of thermalization will in general display various crossovers, it is not clear whether this is a general phenomenon, what are the conditions for its observability and what are going to be its signatures in observables of physical interest.

In this work we consider a Quantum Ising Chain (QIC) perturbed by a time-dependent delta correlated noise in the transverse field direction, and driven out of equilibrium by a quench of the static component of the transverse field. Even though in the last years the non-equilibrium dynamics of a QIC has been studied in great detail theoretically \cite{Rossini,EsslerCalabreseFagotti,Ising}, recently the quench dynamics of a model in the Ising universality class has been realized experimentally in an ensemble of tilted one-dimensional atomic Bose-Hubbard chains \cite{Meinert}, making this problem of potential interest also for experimental studies. Morevoer, recently, the out of equilibrium dynamics of noisy hamiltonians has been studied for trapped bosons and Luttinger liquids  \cite{Noise} and previously, in the framework of open quantum systems, the interplay of many-body interactions, dephasing and dissipation has been studied for spin chains coupled to classical and quantum uniform noise \cite{SpinNoise} or to a bosonic bath \cite{Patane}. As shown by us in a previous work \cite{Marino2012}, the noisy QIC displays \emph{prethermalization} in the time evolution of observables of physical interest (e.g the transverse magnetization).
More specifically, the dynamics in this model has two stages \cite{Marino2012}: first the system relaxes towards the GGE of the unperturbed Ising chain through inhomogeneous dephasing (analogue to dephasing occuring in a Ising chain after the sudden quench of the transverse field); only later noise-induced effects occur, suppressing exponentially fast in time the coherences and subsequently heating the system towards the asymptotic thermal state. The purpose of this paper is to study the system from a complementary point of view, i.e. looking at the statistics of the work done while performing the out of equilibrium protocol discussed above. We consider the probability distribution function of the work done on the system, $P(w)$, (which received an increasing interest in last few years in the domain of quantum quenches \cite{Kehrein}), for static and time-dependent noisy out of equilibrium protocols. We show that in contrast to the noiseless case where a low energy threshold appear with a characteristic edge singularity \cite{Silva}, a sudden quench of the QIC with a static random transverse field drawn from a gaussian distribution function smooths out all non-analyticies in the disorder averaged $P(w)$ (though a definite singularity remains in every realization). On the other hand, the statistics of the work done on an Ising chain with a time dependent noisy magnetic field affects in a time dependent fashion the spectral weight associated to the edge singularity, in a way analogous to what happens in the dynamics of the energy absorbed by the system, presented in \cite{Marino2012} and widely discussed in this work.  This paper is organized as follows. In section II we introduce the model and the out of equlibrium protocol; in section III we start the study of the system, looking at the effect of static and dynamical noise in the work done on a QIC by a noisy protocol. Section IV is devoted to the derivation and the solution of a kinetic equation, using the Keldysh formalism and section V employs these results to study the non-equilibrium dynamics of physical observables in order to understand which are the processes and the time scales involved in thermalization dynamics. Finally, in section VI we summarize our conclusions. Appendix A is devoted to a generalization of Bogolyubov transformations useful for the computation of $P(w)$, when a generic time dependent protocol is performed on the QIC \cite{Smacchia}.\\

\section{The model, the out of equilibrium protocol and the initial state}
The focus of this paper is in the out of equilibrium dynamics of a QIC, described by the hamiltonian
\begin{equation}\label{Hamiltoniana}\begin{split}
H=&H_0+V(t),\\
H_0=&-J\sum_i\sigma^x_i\sigma^x_{i+1}+g\sigma_i^z,\\
V=&\sum_i \delta g(t) \sigma_i^z,\\
\end{split}\end{equation}
where $H_0$ describes the Integrable Quantum Ising Chain and $V(t)$ is a time-dependent gaussian white noise, with zero average and amplitude $\Gamma$,
\begin{equation}\label{delta}\begin{split}
\langle\delta g(t)\rangle&=0,\\
 \langle\delta g(t)\delta g(t')\rangle&=\frac{\Gamma}{2}\delta(t-t').
\end{split}\end{equation}
Here $\widehat{\sigma}_i^{x,z}$ are the longitudinal and transverse spin operators at site $i$ and $g$ is the strenght of the transverse field. The  QIC is among the simplest, yet non-trivial integrable many-body system, whose static properties \cite{Sachdev1999} and quench dynamics \cite{Rossini, EsslerCalabreseFagotti, Ising} are to a great extent known. It is characterized by two dual gapped phases, a quantum paramagnetic ($g>1$) and ferromagnetic one ($g<1$) separated by a quantum critical point located at $g=1$. In the following we assume $J=1$ and we restore it in the computations only when it is necessary.

The spin hamiltonian is unitarily equivalent to spinless fermions, $c_i$, as can be shown by performing a Jordan-Wigner transformation \cite{Sachdev1999}, i.e. defining $\widehat{\sigma}_i^z=1-2c_i^\dag c_i$ and $\widehat{\sigma}_i^+=\prod_{j<i}(1-2c_j^\dag c_j)c_i^\dag$. The Hamiltonian takes in Fourier space, $c_k=\frac{1}{\sqrt{L}}\sum_jc_je^{ikj}$, the simple form
\begin{equation}
H=2\sum_{k>0}\widehat{\psi}_k^\dag \widehat{H_k}\widehat{\psi}_k,
\end{equation}
where
\begin{equation}\label{ham}
 \widehat{H_k}=(g-\cos{k})\sigma_z-(\sin{k})\sigma_y+\delta g(t)\sigma_z
\end{equation}
and $\widehat{\psi}_k$ is the Nambu spinor $\bigl(\begin{smallmatrix}c_k\\ c^\dag_{-k}\end{smallmatrix} \bigr)$ and $\sigma_y$, $\sigma_z$ are the Pauli matrices in the 2$\times$2 Nambu space. The diagonal form $H=\sum_{k>0}E_k(\gamma^{\dag}_k\gamma_k-\gamma_{-k}\gamma^{\dag}_{-k})$, with energies $E_k=\sqrt{(g-\cos k)^2+\sin^2k}$, is achieved after a Bogoliubov rotation $c_k=u_k(g)\gamma_k-iv_k(g)\gamma_{-k}^{\dag}$ and $c_{-k}^{\dag}=u_k(g)\gamma_{-k}^{\dag}-iv_k(g)\gamma_{k}$; the coefficients are given by
\begin{equation}
u_k(g)=\cos(\theta_k(g)) \qquad v_k(g)=\sin(\theta_k(g)),
\end{equation}
where $\tan(2\theta_k(g))=\sin(k)/(g-\cos(k))$. Therefore the QIC can be diagonalized in terms of free fermions, whose mass is the gap of the theory $\Delta=|g-1|$ \cite{Sachdev1999}.

We will consider the dynamics for the following out of equilibrium protocol: at time $t<0$ the system is prepared in the ground state of $H_0$ with a certain value of the transverse field $g_0$, $|\psi_0\rangle=|\psi(g_0)\rangle_{GS}$, and $\delta g(t)=0$. At later time, $t>0$ the system is evolved according to the full hamiltonian $H$ (see \eqref{Hamiltoniana}) with a different value of the transverse field $g$, as portrayed in Fig.1.
\begin{figure}[htbp]
\centering
\includegraphics[scale=0.45]{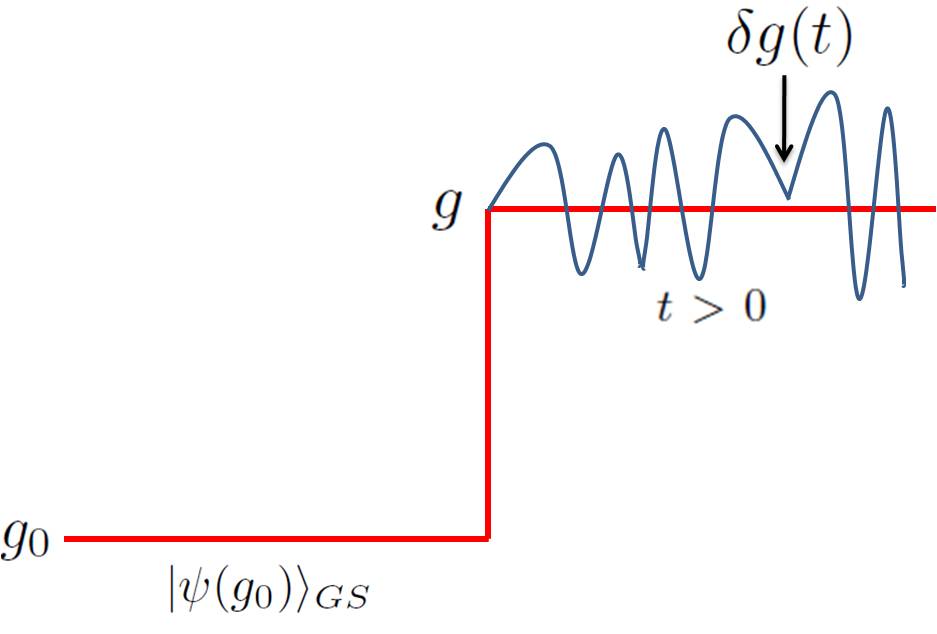}
\caption{[\emph{Colors online}] Out of equilibirum protocol studied in this paper for the QIC: the system is prepared in the ground state of the Ising chain with $g_0>1$ and is evolved according to the Ising Hamiltonian with a different value of the transverse field $g>1$, plus a gaussian delta-correlated noise on top of it. For simplicity, both $g_0$ and $g$ are chosen within the paramagnetic phase.}
\end{figure}

A sudden quench of the transverse field populates all excited states of the system, injecting an extensive ammount of energy; this is easy to understand by looking at the populations and the coherences immediately after the quench. In the basis of the Bogolyubov fermions diagonalizing $H(g)$:

\begin{equation}\label{iniziali}\begin{split}
 \langle\psi_0|\gamma_k^\dag(g)\gamma_k(g)|\psi_0\rangle&=\sin^2(\theta_k-\theta^0_k)\\
 \langle\psi_0|\gamma_k^\dag(g)\gamma_{-k}^\dag(g)|\psi_0\rangle&=-i\frac{\sin 2(\theta_k-\theta^0_k)}{2}\\
 \langle\psi_0|\gamma_{-k}(g)\gamma_k(g)|\psi_0\rangle&=i\frac{\sin 2(\theta_k-\theta^0_k)}{2}\\
 \langle\psi_0|\gamma_{-k}(g)\gamma_{-k}^\dag(g)|\psi_0\rangle&=\cos^2(\theta_k-\theta^0_k),
\end{split}
\end{equation}
where $\theta_k\equiv\theta_k(g)$ and $\theta^0_k\equiv\theta_k(g_0)$. Moreover, the intial state can be written as a coherent superposition of pairs of quasiparticles created on the vacuum of the theory after the quench \cite{Rossini,EsslerCalabreseFagotti} ($H(g)$):
\begin{equation}\label{iniziale}
|\psi(g_0)\rangle_{GS}=N\prod_{k>0}\Big(1+i\tan(\Delta\theta_k)\gamma^\dag_k(g)\gamma^\dag_{-k}(g)\Big)|\psi(g)\rangle_{GS},
\end{equation}

where \begin{equation}\begin{split}
&\Delta\theta_k=\theta_k-\theta^0_k,\\
&N=\exp{\Big[-\frac{1}{2}\sum_{k>0}\log(1+\tan^2\Delta\theta_k)\Big]}.
      \end{split}
\end{equation}

Below we will focus on the interplay between the effect of a sudden quench of $g$ and the time dependent noise driving the dynamics of the system.

\section{Statistics of the work $P(w)$}

The effect of a quantum quench and, generally speaking, of an out-of-equilibrium protocol on a quantum many body system is usually highlighted
studying the time dependence of correlation functions of local operators, as we are going to discuss extensively starting from Section IV. However, since a generic non-equilibrium protocol, as a time dependent magnetic field $g(t)$ in a quantum spin chain, can be seen as the quantum generalization of a thermodynamic
transformation, it could be useful to characterize it studying the work $W$ done on our system upon performing the quench with the noise on the top of it. In a quantum non-equilibrium
process $W$ fluctuates among different realizations of the same protocol \cite{Campisi} and its description requires the introduction of a probability distribution
$P(W)$. On the other hand, work is a fundamental observable in classical and quantum thermodynamics and should be experimentally accessible
by spectroscopic methods, as it has been recently pointed out \cite{Dorner}.

Let us start our analysis by considering the statistic of the work done on a quantum many body system after a quantum quench, $P(w)$ characterized by a generic non-equilibrium protocol $g(t)$. This quantity requires two energy measurements: one at the initial time, $t=\tau_0$, and one at the final time $t=\tau$ (for a comprehensive review on the subject see \cite{Campisi}). We assume that the final energy is measured with respect to the final hamiltonian, $H_{\tau}$,  and that for each realization of the out-of-equilibrium protocol the work $w$ is given as a difference of the outcomes of the two measures of the energy at initial and final time. The statistics of the work is then defined as
\begin{equation}
P(w)=\sum_{n,m}\delta(w-(E_n(\tau)-E_m(\tau_0)))p(n|m,\tau)p_m,
\end{equation}
with $p(n|m,\tau)\equiv|\langle\psi_n(\tau)|U(\tau,\tau_0)|\psi_m(\tau_0)\rangle|^2$, and $p_m\equiv|\langle\psi_m(\tau_0)|\phi(\tau_0)\rangle|$, where $|\phi(\tau_0)\rangle$ is the initial state of the system, $U(\tau,\tau_0)$ is the evolution operator from $\tau_0$ to $\tau$, and $|\psi_i(\tau)\rangle$ are the instantaneous wave-functions, computed from the equation $H_t|\psi_i(t)\rangle=E_i(t)|\psi_i(t)\rangle$. In Ref. \cite{Campisi}, it has been shown that the characteristic function $G(u)=\int dw e^{iuw}P(w)$ contains full information about the statistics of the work $w$ and can be written as a two time correlation function
\begin{equation}\label{char}
 G(u)=\langle e^{iuH^H_{\tau,\tau_0}}e^{-iuH_{\tau_0}}\rangle,
\end{equation}
where $H^H_{\tau,\tau_0}=U^\dag(\tau,\tau_0)H_{\tau,\tau_0}U(\tau,\tau_0)$ is the final Hamiltonian used in the final measurment in the Heisenberg picture. For a sudden quench it follows immediately that
\begin{equation} \label{echo}
G(u)=\langle e^{iH(g_1)u}e^{-iH(g_0)u}\rangle,
\end{equation}
where $H(g_0)$ and $H(g_1)$ are the initial and final hamiltonian respectively.

One may compute exactly the statistics of the work for a generic time variation of the transverse field in the QIC (see Appendix A and Ref. \cite{Smacchia}). For a sudden quench of the transverse field in the QIC, one obtains for $P(w)$, at low $w$, a peak located at $\Delta E_0$, i.e. the difference in the ground states energies before and after the quench, plus a continuum starting above $2\Delta$,  describing pairs of quasi-particles. This continuum displays an edge singularity with universal features \cite{Silva, Smacchia}. For sudden quenches within the paramagentic phase, one may obtain
\begin{equation}\label{stat}
P(\omega)\propto\delta(\omega)+\frac{\sqrt{\pi}}{4}\frac{\Theta(\omega-2\Delta)}{\delta}\rho_{-}^2\sqrt{\frac{\omega-2\Delta}{\Delta}},
\end{equation}
where $\delta=4\pi/L$ is the two-particle level spacing, $\omega=w-\Delta E_0$, $\rho_-=\frac{\Delta_0-\Delta_1}{\Delta_0}$ and $\Theta$ is the Heaviside step function \cite{Silva}. 

Since the exponents of these singularities are expected to be universal \cite{Silva} it is natural to start our study of the effect of the noise by clarifying its role on the universal low-energy behaviour of the statistics of the work. We separate two effects: first we consider a quench with a final random value of the transverse field drawn from a gaussian distribution function and then a gaussian time-dependent delta correlated noise acting on the system during its time evolution till the measurement time $t=\tau$. \\

As a warm up, let us start with the first case, a quench of the QIC with a final value of the transverse field drawn from a gaussian distribution function, corresponding to a value of the final mass, centered in $\overline{\Delta}$ and with variance $\gamma$:
\begin{equation} \label{gauss}
p(\Delta)=\frac{1}{\gamma\sqrt{2\pi}}e^{-\frac{(\Delta-\overline{\Delta})^2}{2\gamma^2}}.
\end{equation}

We now want to compute $P(\omega)$ averaged over this probability distribution. The average energy injected into the system through this quench is equal to the energy injected in a sudden protocol
\begin{equation}\label{inj}\begin{split}
&\Delta E_{injected}\equiv\overline{\langle\psi(g_0)|(H(\overline{g}+\eta)-H(g_0))|\psi(g_0)}=\\
&=\Delta E_{Quench}\equiv\langle\psi(g_0)|(H(\overline{g})-H(g_0))|\psi(g_0)\rangle,
\end{split}\end{equation}
meaning that the noise affects the statistic of the work, $P(\omega)$, starting from the second and higher order moments. Nevertheless, as shown below, the probability distribution (averaged over disorder) is reshaped in the energy window of interest. We can study the statistics of the work by taking the average of \eqref{stat} over the gaussian distribution \eqref{gauss} and assuming $\frac{\gamma}{\overline{\Delta}}\ll1$, $\frac{\gamma}{|\overline{\Delta}-\Delta_0|}\ll1$, i.e. the fluctuations of the noise are small compared to the final gap and the amplitude of the quench.

First of all, it is important to notice that the energy difference of the ground states $\Delta E_0^{noise}$, can be expressed as the difference in the ground states one would have without noise $\Delta E_0$, plus an extensive correction proportional to the fluctuations $\eta=\Delta-\overline{\Delta}$:
\begin{equation}\label{expansion}
\begin{split}
&\Delta E_0^{noise}= E_{GS}(\overline{g}+\eta)-E_{GS}(g_0)\backsimeq \Delta E_0-f(\overline{g})\eta,\\
&\Delta E_0=-\Big(\frac{\overline{g}+1}{\pi}E\Big(\frac{4\overline{g}}{(1+\overline{g})^2}\Big)-\frac{g_0+1}{\pi}E\Big(\frac{4g_0}{(1+g_0)^2}\Big)\Big)
\end{split}
\end{equation}
where we retained only the first order term of the expansion \cite{notaexp} and $E$ is a complete elliptic function. The function $f(\overline{g})$ \cite{nota2} can be expressed as a combination in the following way

\begin{widetext}

\begin{equation}
f(\overline{g})= \frac{L}{\pi}\Big[\frac{1}{\pi}E\Big(\frac{4\overline{g}}{(1+\overline{g})^2}\Big)+\frac{\overline{g}-1}{2(\overline{g}+1)^2}\times  _{2}F_1\Big(\frac{1}{2},\frac{3}{2},2,\frac{4\overline{g}}{(\overline{g}+1)^2}\Big)\Big]\equiv L\varXi(\overline{g}).
\end{equation}

\end{widetext}

where $_{2}F_1$ is an hypergeometric function.

Below we focus on the \emph{average} statistics of the work

\begin{widetext}

\begin{equation}\label{averagePW}\begin{split}
P(\omega)\propto& \int_{-\infty}^{\infty}d\eta \frac{e^{-\frac{\eta^2}{2\gamma^2}}}{\sqrt{2\pi\gamma}} \Big[ \delta(\omega+f(\overline{g})\eta)+
\frac{\sqrt{\pi}}{4}\frac{\Theta(\omega+f(\overline{g})\eta-2\overline{\Delta}-2\eta)}{\delta}\Big(\frac{\Delta_0-\overline{\Delta}-\eta}{\Delta_0}\Big)^2\sqrt{\frac{\omega+f(\overline{g})\eta-2\overline{\Delta}-2\eta}{\overline{\Delta}+\eta}} \Big]\simeq \\
&\simeq \frac{e^{-\frac{\omega^2}{2(\gamma f(\overline{g}))^2}}}{\sqrt{2\pi}\gamma f(\overline{g})}+ \int_{-\infty}^{\infty}d\eta \frac{e^{-\frac{\eta^2}{2\gamma^2}}}{\sqrt{2\pi\gamma}}\frac{\sqrt{\pi}}{4}\frac{\Theta(\omega+f(\overline{g})\eta-2\overline{\Delta}-2\eta)}{\delta}\rho_-^2\sqrt{\frac{\omega+f(\overline{g})\eta-2\overline{\Delta}-2\eta}{\overline{\Delta}}}
\end{split},
\end{equation}

\end{widetext}

where in the second line we assumed $\frac{\gamma}{\overline{\Delta}}\ll1$ and $\frac{\gamma}{|\overline{\Delta}-\Delta_0|}\ll1$.
This formula contains two physical effects, the first one is a global fluctuation involving the shift of the ground state energy (see Eq. \eqref{expansion}). This effect is proportional to the system size $L$ and affects in the same way both the delta peak singularity and the continuum starting at $\omega=2\Delta$. The second effect is associated to the fluctuations affecting the masses of the quasi-particles emitted after the quench and it does not scale with the size of the system. If one is interested in measuring the work with reference to $\Delta E_0$ in an energy window close to $\Delta E_0+2\overline{\Delta}$, the first type of fluctuations are obviously dominant and most importantly detrimental. Indeed, the last integral in Eq. \eqref{averagePW} can be cast in the following form $A\sqrt{\gamma'}\int_{-c}^\infty dy e^{-y^2/2}\sqrt{y+c}$ (where $A=\frac{1}{4\sqrt{2}}\frac{1}{\delta}\frac{\rho_-^2}{\sqrt{\overline{\Delta}}}$, $\gamma'=\gamma(f(g)-2)$ and $c=\frac{\omega-2\overline{\Delta}}{\gamma'}$). At energies around $2\overline{\Delta}$ one would observe

\begin{equation}
\overline{P(\omega)}\simeq \frac{1}{\sqrt{2\pi}\gamma'} + C\frac{\rho_-^2}{{\delta}}\sqrt{\frac{\gamma'}{\overline{\Delta}}}\Big(\frac{1}{\Gamma(\frac{5}{4})}+\frac{\sqrt{2}}{\Gamma(\frac{3}{4})}\frac{\omega-2\overline{\Delta}}{\gamma'}+...\Big)
\end{equation}

where $C$ is a numerical prefactor and $\Gamma$ is the Euler Gamma function.

It could be interesting to subtract these fluctuations by some means. In order to to so there are in principle two possibilities: the first one is to measure for each realization only the energy differences with respect to the threshold, subtracting the extensive shift of the ground state energy due to the noise (see Eq.\eqref{expansion}); the second one consists in rescaling the noise amplitude by the system size, $\gamma\rightarrow\frac{\gamma}{L}$. In both ways Eq. \eqref{averagePW} can be properly averaged in the energy range of interest. 
For $\omega-2\overline{\Delta}\gg\gamma'$
\begin{equation}
P(\omega)\propto P_{quench}(\omega)\Big(1+O\Big(\frac{\gamma\varXi(g)}{(\omega-2\overline{\Delta})}\Big)^2\Big),
\end{equation}
which essentially means that well above the energy threshold for the production of pairs of quasi-particles in a sudden quench, the statistics of the work is left unchanged. On the other hand, for $\omega\ll2\overline{\Delta}-\gamma'$, the statistics of the work displays a gaussian tail controlled by the renormalized noise amplitude $\gamma'$,
\begin{equation}\label{coda}
P(\omega)\propto\frac{\rho_-^2}{\delta}\sqrt{\frac{\gamma'}{\overline{\Delta}}}\Big(\frac{\gamma' }{|\omega-2\overline{\Delta}|}\Big)^{3/2}e^{-\frac{(\omega-2\overline{\Delta})^2}{2\gamma'^2}}.
\end{equation}
\\

Let us now proceed our analysis considering more complicated effects. We prepare the system in the ground state of the Ising chain in the paramagnetic phase, with $g_0>1$ and we let evolve the system under the generic time-dependent hamiltonian $H_0+V(t)$. In the following we assume that we have subtracted the shift of the ground state energy and that the amplitude of the noise has been rescaled.

It is a remarkable fact that for each realization of the noise the square root singularity at the lower energy threshold is independent from the out-of equilibrium protocol performed on the QIC \cite{Smacchia}; what changes is the spectral weight of the singularity in $P(w)$, which in general will depend on the details of the time dependent quench, as discussed in Appendix A. The expression of the statistics of the work in this case is
\begin{equation}\label{statT}
P(\omega, \tau)\simeq\delta(\omega)+\frac{\sqrt{\pi}}{4}\frac{\Theta(\omega-2\Delta(\tau))}{\delta}|\rho(\tau)|^2\sqrt{\frac{\omega-2\Delta(\tau)}{\Delta(\tau)}},
\end{equation}
where
\begin{equation}\label{rhodipT}\begin{split}
&|\rho(\tau)|^2\equiv\Delta^2(\tau)\Big|\rho- \int_0^\tau\frac{e^{2i\int_0^tdt'\Delta(t')}}{\Delta(t)^2}\dot{\Delta}(t)dt\Big|^2 \\
&= \Delta^2\Big(\rho^2-2\rho Re\Big[\int_0^\tau\frac{e^{2i\int_0^tdt'\Delta(t')}}{\Delta(t)^2}\dot{\Delta}(t)dt\Big]+\\
&+\Big|\int_0^\tau\frac{e^{2i\int_0^tdt'\Delta(t')}}{\Delta(t)^2}\dot{\Delta}(t)dt\Big|^2\Big)
\end{split}\end{equation}
and $\rho=\frac{\Delta_0-\Delta(0)}{\Delta_0\Delta(0)}$, where in general $\Delta(0)$ is different from $\Delta_0$.

The derivation of Eq. \eqref{statT} is postponed in Appendix A. Using integration by parts, it is easy to show that
\begin{equation}\label{parti}\begin{split}
&\int_0^\tau\frac{e^{2i\int_0^t dt'\Delta(t')}}{\Delta(t)^2}\dot{\Delta}(t)dt=\\
&=\frac{1}{\Delta(0)}-\frac{1}{\Delta(\tau)}e^{2i\int_0^\tau dt'\Delta(t')}
+2i\int_0^\tau dt e^{2i\int_0^tdt'\Delta(t')},
\end{split}\end{equation}

When taking the noise average of these expressions there are going to be two separate effects. The first will consist in fluctuations of $\Delta$ at the initial and final point of the trajectory which will produce consequences similar to the ones discussed above in the static case. If we think to the statistics of $\Delta(t)$ as being Gaussian with:
\begin{equation}\label{rumore}
\langle\Delta(t)\Delta(t')\rangle\simeq\frac{\Gamma}{2}\delta_{\tau_c}(t-t'),
\end{equation}
where $\tau_c$ is a correlation time \cite{nota7}, the fluctuations at the endpoints have amplitude $\gamma=\sqrt{\frac{\Gamma}{\tau_c}}$. Now in the limit, $\frac{\gamma}{\Delta}$, $\frac{\gamma}{|\Delta-\Delta_0|}\ll1$ we argue that to the leading order the various terms in Eq. \eqref{statT} can be averaged separately:
\begin{equation}
P(\omega, \tau)\simeq\delta(\omega)+\frac{\sqrt{\pi}}{4}\frac{\Theta(\omega-2\Delta(\tau))}{\delta}\overline{|\rho(\tau)|^2}\overline{\sqrt{\frac{\omega-2\Delta(\tau)}{\Delta(\tau)}}}.
\end{equation}
While the average of the square root singularity will produce the smearing of the singularity described above, the average of the spectral weight will produce a time dependent prefactor that appears to describe the heating of the system under the influence of the noise.
In order to average $\overline{|\rho(\tau)|^2}$, we first notice that for $\frac{\gamma}{\Delta}$, $\frac{\Gamma}{\Delta}\ll1$, we have
\begin{equation}
\overline{\frac{1}{\Delta(\tau)}e^{2i\int_0^\tau dt'\Delta(t')}}\simeq\frac{1}{\overline{\Delta(\tau)}}\overline{e^{2i\int_0^\tau dt'\Delta(t')}}\simeq\frac{1}{\Delta}e^{-\Gamma\tau} e^{2i\Delta\tau},
\end{equation}
where crossed correlations with the boundary term proportional to $\Delta(\tau)$ can be neglected. Indeed, expanding in Taylor series the left hand side, we get
\begin{equation}\begin{split}
&\frac{e^{2i\Delta\tau}}{\Delta}\times\Big(1-\frac{\eta(\tau)}{\Delta}+\frac{\eta(\tau)^2}{\Delta^2}+...\Big)\times\\
&\times\Big(1+2i\int_0^\tau dt'\eta(t')+\frac{1}{2}(2i)^2\int_0^\tau dt'dt''\eta(t')\eta(t'')+...\Big)
\end{split}\end{equation}
and, taking the average over the noise, we finally have
\begin{equation}\label{serietaylor}
\begin{split}
&\frac{e^{2i\Delta\tau}}{\Delta}e^{-\Gamma\tau}\times\\
&\times\Big(1-i\frac{\Gamma}{\Delta}+
\Big(\frac{\gamma}{\Delta}\Big)^2-\Big(\frac{\Gamma}{\Delta}\Big)^2-i\Big(\frac{\gamma}{\Delta}\Big)^2\frac{\Gamma}{\Delta}+i\Big(\frac{\Gamma}{\Delta}\Big)^3+...\Big)
\end{split}
\end{equation}
It should be clear that in the limit $\frac{\gamma}{\Delta}\ll1$, $\frac{\Gamma}{\Delta}\ll1$, only the first term can be kept in the right hand side of \eqref{serietaylor}.\\


Using Eq. \eqref{parti}, Eq. \eqref{rumore}, and neglecting correlations coming from boundary terms, it is now straightforward to average over the noise; for instance, for the second term in Eq. \eqref{rhodipT} we get
\begin{widetext}
\begin{equation}\label{secondo}\begin{split}
&\overline{Re\Big[\int_0^\tau\frac{e^{2i\int_0^tdt'\Delta(t')}}{\Delta(t)^2}\dot{\Delta(t)}dt\Big]}=\frac{1}{\Delta}\Big(1-e^{-2\Gamma\tau}\cos(2\Delta\tau)\Big)+\frac{1}{\Delta^2+\Gamma^2}\Big[\Delta\Big(e^{-2\Gamma\tau}\cos(2\Delta\tau)-1\Big)+\Gamma e^{-2\Gamma\tau}\sin(2\Delta\tau)\Big]\simeq \\
&\substack{\ \\[1mm]\simeq\\\frac{\Gamma}{\Delta}\ll1}\frac{\Gamma}{\Delta^2}e^{-2\Gamma\tau}\sin(2\Delta\tau).
\end{split}\end{equation}
which is of order $\frac{\Gamma}{\Delta}$ when reinserted in \eqref{rhodipT}.\\

The third contribution can be written as
\begin{equation}\label{terzo}\begin{split}
&\Big|\int_0^\tau\frac{e^{2i\int_0^tdt'\Delta(t')}}{\Delta(t)^2}\dot{\Delta}(t)dt\Big|^2\equiv\\
&\equiv\Big(\frac{1}{\Delta(0)}-\frac{1}{\Delta(\tau)}e^{2i\int_0^\tau dt'\Delta(t')}
+2i\int_0^\tau dt e^{2i\int_0^tdt'\Delta(t')}\Big)\times
\Big(\frac{1}{\Delta(0)}-\frac{1}{\Delta(\tau)}e^{-2i\int_0^\tau dt'\Delta(t')}
-2i\int_0^\tau dt e^{-2i\int_0^tdt'\Delta(t')}\Big)=\\
&=\Big|\frac{1}{\Delta(0)}-\frac{1}{\Delta(\tau)}e^{2i\int_0^\tau dt'\Delta(t')}\Big|^2+2Re\Big[2i\Big(\frac{1}{\Delta(0)}-\frac{1}{\Delta(\tau)}e^{-2i\int_0^\tau dt'\Delta(t')}\Big)\int_0^\tau dt e^{2i\int_0^t dt'\Delta(t')}\Big]+\\
&+4\int_0^\tau dt e^{2i\int_0^tdt'\Delta(t')}\times\int_0^\tau dt e^{2i\int_0^tdt'\Delta(t')}.
\end{split}\end{equation}
\end{widetext}

Under the same approximations stated above and using again \eqref{parti}, it is possible to average \eqref{terzo} over the time-dependent noise \eqref{rumore}, disregarding noise fluctuations in the boundary terms proprotional to $\Delta(0)$ and $\Delta(\tau)$. To compute the average of \eqref{terzo}, we need to average products of two noise dependent quantities; for instance, it is easy to derive
\begin{equation}\begin{split}
&\overline{4\int_0^\tau dt e^{2i\int_0^tdt'\Delta(t')}\times\int_0^\tau dt e^{2i\int_0^tdt'\Delta(t')}}\simeq\\
&\substack{\ \\[1mm]\simeq\\\frac{\Gamma}{\Delta}\ll1}4\frac{\Gamma\tau}{\Delta^2}+O\Big(\frac{\Gamma}{\Delta}\Big)
\end{split}\end{equation}
while all the other terms in \eqref{terzo} are subleading in the limit $\frac{\Gamma}{\Delta}\ll1$ and $\frac{\gamma}{\Delta}\ll1$.\\

Hence, our result on the statistics of the work, $P(\omega,\tau)$, can be summarized in the following expression which contains a transparent physical meaning

\begin{equation}
P(\omega,\tau)\simeq\delta(\omega)+(\rho_-^2+4\Gamma\tau)Q(\omega)
\end{equation}
where
\begin{equation}
Q(\omega)=\frac{\sqrt{\pi}}{4}\overline{\frac{\Theta(\omega-2\Delta)}{\delta}\sqrt{\frac{\omega-2\Delta}{\Delta}}}.
\end{equation}

The long time growth of the spectral weight appears to indicate the continous heating of the system (it resembles the time dependence of the energy absorbed by the system at the early stages of the dynamics, as it will be clear from Eq.\eqref{EnIniz}). Notice indeed that the energy absorbed by the system during the time-dependent protocol $g(t)$ is non zero, in sharp contrast to the static case, as we will show in Section V. In the following we will study in more sophisticated quantities the interplay between dynamical noise and coherent effects due to a quantum quench of the Ising Chain.

\section{Kinetic equations}

In this section we are going to study the kinetics of local observables and their correlation functions in the QIC. In order to accomplish this task, we are interested in deriving a kinetic equation for the equal time non-equilibrium Green's function for the protocol discussed in Section II. We will do so by deriving a master equation, using the Keldysh contour technique, in order to obtain analitically an expression for the 2-point functions of Bogolyubov fermions at equal time. These equations will then be used to compute all the observables of interest and their the out-of-equilibrium dynamics. Part of the results presented in this section have been announced in Ref.\cite{Marino2012}.\\

\begin{widetext}

\begin{figure}[htbp]
\centering
\includegraphics[scale=0.55]{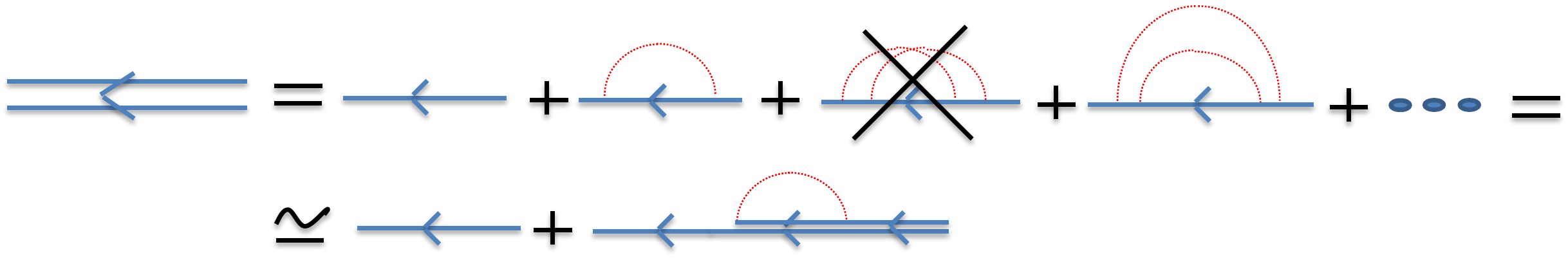}
\caption{A diagramatic representation of the Dyson series \eqref{dyson}. Crossed dyagrams are neglected according to the \emph{self-consistent Born approximation}.}\label{dys}
\end{figure}

\end{widetext}

We start recalling the definition of the statistical Green function on the Keldysh contour  \cite{Haug}
\begin{equation}
 G^c=-i\langle T_c\psi_{ki}(\tau)\psi^\dag_{kj}(\tau')\rangle,
\end{equation}
where $T_c$ is the time ordering operator on the Keldysh contour, $\tau$ and $i$ and $j$ are indices in the Nambu space; we define the lesser Green function as
\begin{equation}
 G^<(t,t')=\Big[G_k^<(t,t')\Big]_{i,j}=i\langle\psi_{k,j}^\dag(t')\psi_{k,i}(t)\rangle,
\end{equation}
which is a matrix in the Nambu space (here $t$ and $t'$ are real times).\\
Using the standard approach \cite{Haug}, we first write the equation for the statistical Green function with the noise as a perturbation and we resum the Dyson series (Fig.~\ref{dys})
\begin{equation}\label{dyson}
 G^c_{\tau,\tau'}=G_{0_{\tau,\tau'}}^c+G_{0_{\tau,\tau''}}^c\otimes\Sigma^c_{\tau'',\tau'''}\otimes G^c_{\tau''',\tau'}
\end{equation}
where $G_{0_{\tau,\tau'}}^c$ is the unperturbed Green function and $\Sigma^c_{\tau,\tau'}$ is the self energy;  in right hand side the simbol $\otimes$ is understood as a convolution product, all the quantities are evaluated along the Keldysh contour.

In the followig we will neglect noise crossed diagrams, computing the self-energy within the so called \emph{self-consistent Born approximation} \cite{Haug}, controlled by the small parameter $\frac{\Gamma}{\Delta}$, as illustrated in Fig. 2. This dimensionless parameter is, in a sense, the
analogue of $k_Fl\gg1$ in disordered electron systems, where the typical length scale associated to electron wavefunctions, $\lambda_F\sim1/k_F$ ($k_F$ is the Fermi wave vector), is much smaller than the typical length associated to disorder, $l$ (the average mean path), and correlations induced by the latter can be disregarded at leading order in $k_Fl\gg1$. This physical analogy is at the origin of the approximation $\frac{\Gamma}{\Delta}\ll1$, since $\Gamma$ and $\Delta$ can be considered the analogue of $\frac{1}{l}$ and $k_F$ respectively.

The Dyson equation for the statistical Green function is then
\begin{equation}\label{dyson}
\begin{split}
 i\partial_tG^<(t,t')=&H_kG^<(t,t')+\int dt'' [\Sigma^<(t,t'')G^a(t'',t')+\\
& +\Sigma^r(t,t'')G^<(t'',t')],\\
 -i\partial_{t'}G^<(t,t')=&G^<(t,t')H_k+\int dt'' [G^r(t,t'')\Sigma^<(t'',t')+\\
& +G^<(t,t'')\Sigma^a(t'',t')].
\end{split}
\end{equation}
Within the self consistent Born approximation, we obtain for the self energies in \eqref{dyson}:
\begin{equation}\label{sigma}\begin{split}
 \Sigma^<_{t,t'}&=\frac{\Gamma}{2}\delta(t-t')\sigma_zG^<_{t,t'}\sigma_z \\
 \Sigma^{r,a}_{t,t'}&=\mp i\frac{\Gamma}{4}\delta(t-t').
\end{split}
\end{equation}
We substitute \eqref{sigma} in \eqref{dyson}, subtract the two resulting equations and take the limit $t\rightarrow t'$; defining the density matrix
\begin{equation}
 \rho_k(t)=-iG^<_k(t,t)
\end{equation}
we finally obtain the master equation
\begin{equation}\label{master}
\delta_t\rho_k=-i[H_k,\rho_k]+\frac{\Gamma}{2}(\sigma_z\rho_k\sigma_z-\rho_k),
\end{equation}
where $[H_k,\rho_k]$ is responsible for the free dynamics and the second term on the right hand side contain information about the dissipation due to the noise.
We now apply to \eqref{master} a Bogolyubov rotation $U(\theta_k)=\exp(-i\theta_k \sigma_{x})$ with $\theta_k=1/2\arctan[(\sin k)/(g-\cos k)]$, which diagonalizes the Ising model in the basis of the Bogoliubov fermions $\gamma_k$.
We get
\begin{equation}\label{master2}
\partial_t\rho_k=-i[\widetilde{H_k},\rho_k]+\frac{\Gamma}{2}(\sigma'\rho_k \sigma'-\rho_k),
\end{equation}
where $\sigma'=U^\dag(\theta_k)\sigma_zU(\theta_k)=\cos2\theta_k\sigma_z+\sin2\theta_k\sigma_y
$ and the density matrix is expressed in the basis of the Bogoliubov fermions.

Before solving Eq. \eqref{master2}, let us comment on the properties of the noise. In the base diagonalizing the final hamiltonian, $H_k$ appears as
\begin{equation}\label{diagonale}\begin{split}
 H_k=&E_k\sigma_z+\delta g(t)(\sigma_z\cos 2\theta_k+\sigma_y\sin 2\theta_k)=\\
&E_k\sigma_z+\delta g_k^z(t)\sigma_z+\delta g_k^y(t)\sigma_y,
\end{split}\end{equation}
where $\delta g_k^z(t)$ and $\delta g_k^y(t)$ statisfy
\begin{equation}\begin{split}
\langle\delta g_k^z(t)\delta g_k^z(t')\rangle&=\frac{\Gamma}{2}(\cos 2\theta_k)^2\delta(t-t'),\\
\langle\delta g_k^y(t)\delta g_k^y(t')\rangle&=\frac{\Gamma}{2}(\sin 2\theta_k)^2\delta(t-t'),\\
\end{split}\end{equation}
where it should be easy to see that our model is equivalent to the QIC perturbed by two $k$-dependent delta correlated noises, one along the $z$ direction and the other one along $y$. Morevoer the noise along the $y$ direction is correlated to the noise along the $z$ direction
\begin{equation}
\langle\delta g_k^z(t)\delta g_k^y(t')\rangle=\frac{\Gamma}{2}\sin 2\theta_k \cos 2\theta_k\delta(t-t').
\end{equation}
The usual way to solve a master equation like \eqref{master2} is to decompose the density matrix in the basis of the Pauli matrices
\begin{equation}\label{ansatz}
\rho_k=\frac{1}{2}\textbf{1}+\delta f_k\sigma_z+x_k\sigma_x+y_k\sigma_y.
\end{equation}
Plugging this decomposition in the master equation \eqref{master2} we end up with a system of differential equations for the coefficients of the density matrix \eqref{ansatz}
\begin{equation}\begin{split}\label{sistema}
\partial_t(\delta f_k)&=-\Gamma\sin^22\theta_k\delta f_k+\frac{\Gamma}{2}y_k\sin4\theta_k\\
\partial_t x_k&=-\Gamma x_k-2E_ky_k\\
\partial_t y_k&=\frac{\Gamma}{2}\sin4\theta_k\delta f_K+2E_kx_k-\Gamma\cos^22\theta_k y_k.
\end{split}
\end{equation}
We will in the following solve this system of equations in the limit $\frac{\Gamma}{\Delta}\ll1$, which allows to neglect $y$-$z$ correlations; we checked this approximation  numerically for different values of $k$ in the Brillouin zone. Taking into account the different initial conditions \eqref{iniziali}, corresponding to an extensive amount of energy injected in the system by the quench of the transverse field, we immediately obtain
\begin{equation}\label{esatta}
\delta f_k(t)=(\sin^2(\Delta\theta_k)-1/2) e^{-\Gamma t\sin^22\theta_k}.
\end{equation}

For the coherences $z_k=x_k-iy_k$ we instead obtain
\begin{equation}\label{zetatot}
\partial_t z_k=(2E_ki-\Gamma)z_k+\frac{\Gamma}{2}(1-\cos^2(2\theta_k))\frac{z_k-z_k^*}{2};
\end{equation}
from this equation we see that the coherences decay exponentially fast as $\Gamma t\gg1$, as one can see close to $k\simeq0,\pi$:
\begin{equation}\label{zeta}
z_k\simeq z^0_ke^{2iE_kt}e^{-\Gamma t}.
\end{equation}
On the other hand, from equation \eqref{esatta}, we see that while most of the modes relax fast to their thermal occupation ($n_k \simeq 1/2$) on time scales of the order of $1/\Gamma$, the relaxation rates tend to vanish close to the band edges ($k=0, \pm \pi$) (see Fig. 3).




We give the expression for $\delta f_k$ and $z_k$ for $k\simeq0$, as they will be useful to compute the leading behaviour of physical observables during thermalization dynamics, as it will be more clear in the next sections:

\begin{equation}\label{decadimento}
\begin{split}
\langle\gamma_k^\dag\gamma_k\rangle&=\frac{1}{2}+\frac{1}{2}\Big(\frac{k^2}{2\Delta^2}\rho_{-}^2-1\Big)e^{\frac{-\Gamma k^2t}{\Delta^2}}\\
\langle\gamma_k^\dag\gamma_{-k}^\dag\rangle&=-\frac{ik}{2\Delta}\rho_{-}e^{-\alpha t-i\beta t},
\end{split}
\end{equation}
 where $\rho_{-}\equiv\frac{\Delta_0-\Delta}{\Delta_0}$ and \begin{equation}\begin{split}
 \alpha=&\Gamma\Big(1-\frac{1}{2}\Big(\frac{k}{\Delta}\Big)^2\Big)\\
 \beta=&2\Delta\Big(1+\frac{1}{2}\Big(\frac{k}{\Delta}\Big)^2\Big).
\end{split}
\end{equation}


\begin{figure}[htbp]
\centering
\includegraphics[scale=0.8]{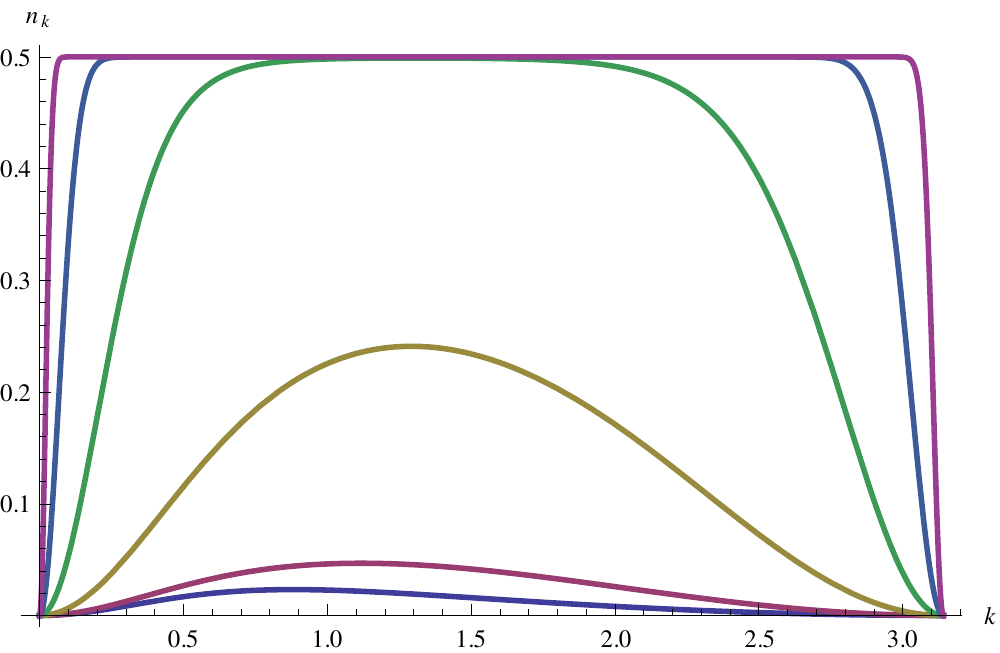}
\caption{Populations, $n_k=\langle\gamma_k^\dag\gamma_k\rangle$ \emph{vs} wave vector $k$ at different times: from bottom to up, $\Gamma t=0.1, 1, 10, 10^2, 10^3, 10^4$ ($g_0=2$, $g=4$).}
\end{figure}



\section{Thermalization Dynamics of Observables}

Let us start now the study of the interplay between quench and noise in the time evolution of observables of interest, studying their dynamics from the intial state towards the asymptotic steady state, which is the infinte temperature state, where all fermion modes are equally occupied, $n_k=1/2$, for all $k$ in the Brillouin zone. We shall start computing the energy absorbed by the system. We will then be concerned with the study of thermalization dynamics of the transverse magnetization correlator and, finally, we are going to look for signatures of the noise in the time evolution of the order parameter correlations.

\subsection{Energy absorbed by the QIC}

Let us start considering the energy absorbed by the system during the noisy time-dependent protocol:
\begin{equation}
E(t)= \langle\psi(t)|H(g(t))|\psi(t)\rangle,
\end{equation}
where $|\psi(t)\rangle$ is the state at time $t$. Substituting the expression for the hamiltonian \eqref{ham}, we get
\begin{equation}\begin{split}
E(t)=&\langle \psi(t)|\Big(H_0(g)+\delta g(t)\sum_i\sigma_i^z\Big)|\psi(t)\rangle=\\
=&\langle\psi(t)|H_0(g)|\psi(t)\rangle+\delta g(t)\langle\psi(t)|\sum_i\sigma_i^z|\psi(t)\rangle.
\end{split}
\end{equation}
Let us now assume that at the time $\tau$ and onwards the noise is turned off. Therefore the total energy acquired at time $\tau$ by the system is
\begin{equation}\label{energia}
E(\tau)=N\int_0^\pi \frac{dk}{2\pi} E_k(g)(\langle\gamma_k^\dag(\tau)\gamma_k(\tau)\rangle-\langle\gamma_{-k}(\tau)\gamma_{-k}^\dag(\tau)\rangle).
\end{equation}
We can now use the expectation values for the two-point functions of the Bogolyubov fermions derived in the previous section to evaluate this expression as a function of $\tau$. For times $\Gamma \tau\ll1$, the energy is equal to the energy injected in an ordinary quench $E_{Quench}$ plus small corrections
\begin{equation}\label{EnIniz}
E(\tau)=E_{Quench}+N\int_0^\pi\frac{dk}{2\pi}\epsilon_k\cos(2\Delta\theta_k)\sin^2(2\theta_k) \Gamma\tau,
\end{equation}
where $E_{Quench}=-\frac{N}{2\pi}\int_0^\pi dk\epsilon_k\cos(2\Delta\theta_k)$ is the energy injected in the system by a sudden quench.

At longer times, $\Gamma t\gg1$, the energy saturates towards its asympotic limit, zero with our choice of the vacuum energy, with an asymptotic power law behaviour $\frac{1}{\sqrt{\Gamma t}}$, which is the signature of the slow relaxation of  $k\simeq0,\pi$ modes, discussed in Section IV \cite{nota3}. In particular, \eqref{energia} can be written as \begin{equation}\begin{split}
E(t)&=\frac{N}{2\pi}\int_0^\pi2E_k\delta f_k=\\
&=-\frac{N}{2\pi}\int_0^\pi dk E_k \cos{2\Delta \theta_k}e^{-\Gamma t\sin^22\theta_k}
\end{split}\end{equation}
and for $\Gamma t\gg1$ this quantity is dominated by the modes with smallest relaxation rate, $k\simeq0, \pi$, with the final result
\begin{equation} E(t)\substack{\ \\[1mm]\simeq\\ \Gamma t\gg1}-\frac{N}{2\sqrt{\pi}}\frac{g^2+1}{\sqrt{\Gamma t}}.\end{equation}

\subsection{Evolution of the number of kinks}

Let us now turn our attention to a more interesting quantity to highlight the dynamics of thermalization: the density of the number of kinks, defined as
\begin{equation}
n_{kink}\equiv\frac{1}{2N}\sum_i\langle(1-\sigma_i^x\sigma_{i+1}^x)\rangle.
\end{equation}
Simple algebraic manipulations yield
\begin{equation}\label{kink}\begin{split}
n_{kink}&(t)=\frac{1}{2N}\sum_{k}(1+2\langle\gamma_k^{\dag}(g=0)\gamma_k(g=0)\rangle)=\\ =&\frac{1}{2N}\sum_{k}\Big(2+2\delta f_k(t)\cos2\Delta\alpha_k^*+2y_k(t)\sin2\Delta\alpha_k^*\Big).
\end{split}\end{equation}
This result has been obtained by expressing Bogoliubov fermions at $g=0$ in terms of fermions diagonalizing the chain at finite $g$, consequently $\Delta\alpha_k^*=\theta_k(g=0)-\theta_k(g)$ is the difference between the two angles.
It is clear from this expression that the number of kinks can be written as the sum of two terms, $n_{kink}(t)\equiv n_{drift}(t)+\Delta n(t)$, the first due to populations (plus the constant term) and describing the heating of the system towards the asympotic steady state and the second one responsibile for dephasing and exclusively due to coherences, which is at the origin of an intermediate stage of the dynamics of $n_{kink}$, which we shall relate to \emph{prethermalization}. \\
Thermalization dynamics of $n_{kink}(t)$ can be divided in three stages as summarized in Fig. 4:
\begin{figure}[htbp]
\centering
\includegraphics[scale=0.8]{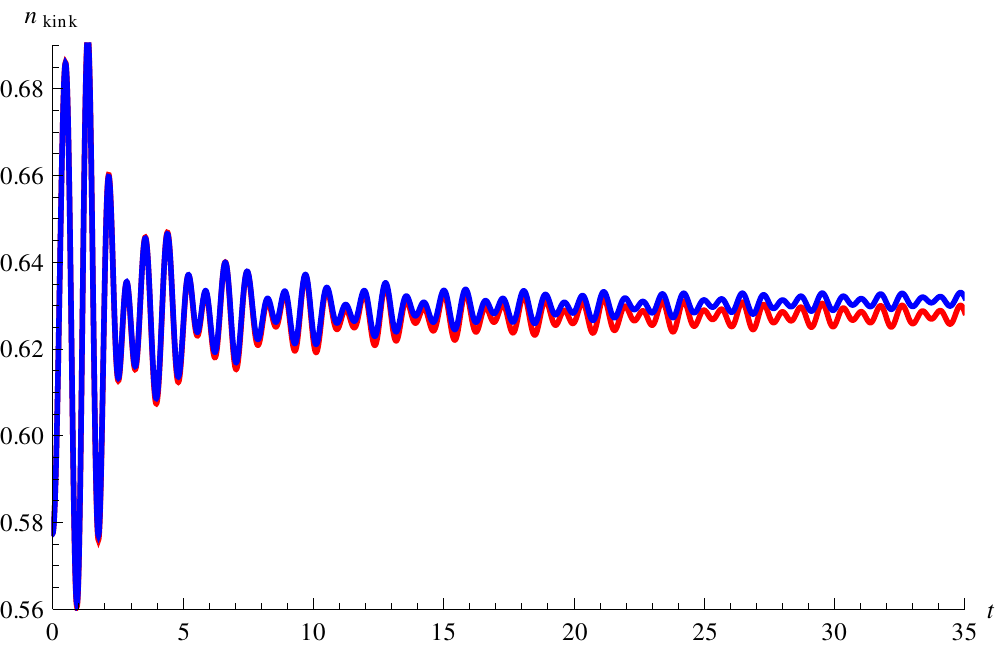}
\caption{[\it Colors online \rm]  The density of kinks vs. time for a quench  with $\Gamma=0.01$, $g_0=1.1$, $g=4$. While the red line shows the value attained by $n_{\rm kink}$ without perturbation and predicted by the GGE,  the full time evolution (blue line) shows first a saturation towards the GGE value and later a runaway towards the infinite temperature state.}\label{kinks}
\end{figure}
\begin{enumerate}
 \item first of all, the system relaxes towards the asymptotic steady state of the QIC after a quench of the transverse field without noise, which is the GGE of the QIC, accounting for the conserved quantities of the theory, i.e. the occupation number of the fermions $n_k=\gamma_k^\dag\gamma_k$. This happens through the usual inhomogeneous dephasing \cite{Ising}, arising from the overlap of a continuum of frequencies in \eqref{kink} and leading to a $\frac{1}{(Jt)^{3/2}}$ decay in the $Jt\gg1$ limit. This result can be easily derived applying a stationary phase argument to Eq. \eqref{kink} in the $Jt\gg1$ limit and in the temporal frame when the noise is not effective $\Gamma t\ll1$. Though the term \emph{prethermalization} has been introduced for closed quantum many body systems driven out of equilibrium, the appearence of an intermediate stage of the dynamics observed here is very similar to what have been found in closed systems \cite{Kollar2011}, suggesting to use this term also in this context.
\item The second stage consists in a noise induced dephasing, where coherences are suppressed exponentially by the noise for $\Gamma t\gg1$, as the leading $e^{-\Gamma t}$ behaviour discussed before suggests.
\item The third stage corresponds to populations heating up. This drives the number of kinks towards the final stage of the dynamics, i.e an infinte temperature state. This happens following the same $\frac{1}{\sqrt{\Gamma t}}$ behaviour of the energy, and it is due again to the presence of slow relaxing modes dominating thermalization dynamics.
\end{enumerate}
This scenario can be better understood by looking separately at $n_{drift}$ and $\Delta n_{kink}$. In Fig.5, $n_{drift}$ is plotted as a function of time (red line), showing that this term is responsible for the deviation of $n_{kink}$ from the GGE expectation value (blue line), while $\Delta n_{kink}$, plotted in Fig.6, first decays following a power law, while for times $\Gamma t\gg1$ it starts decaying exponentially fast, departing clearly from the values attained in the usual sudden quench protocol (blue line).\\

As a last remark in this Section, it should be noticed that the appearence of
{prethermalization} stage strictly depends on the different behaviour of the populations and coherences during the dynamics. This implies that whether an observable will show \emph{prethermalization} or not will depend crucially on its expression in the Bogolyubov basis. This is the reason beneath the absence of a similar behaviour in the dynamics of $E(t)$.\\

\subsection{On-site transverse magnetization}

A \emph{pre-thermal} plateau would be also observed in the thermalization dynamics of the on-site transverse magnetization, $\langle\sigma_i^z(t)\rangle$, which posses a similar expression to \eqref{kink} in the Bogolyubov basis
\begin{equation}
m^z\equiv\langle\sigma_i^z\rangle=\int_0^\pi dk\frac{2}{\pi}\Big(\delta f_k(t)\cos2\theta_k-\sin2\theta_ky_k(t)\Big).
\end{equation}
The pre-thermal plateau is in correspondence of the expectation value of $\sigma_i^z$ evaluated in the GGE of the QIC without noise
\begin{equation}
\langle\sigma_i^z\rangle_{GGE}=-\int_0^\pi dk\frac{1}{\pi}\cos2\Delta\theta_k\cos2\theta_k
\end{equation}
and it is approached with a power law, $\frac{1}{(Jt)^{3/2}}$, in the limit $Jt\gg1$, as in a quenched QIC \cite{EsslerCalabreseFagotti}. On the other hand, the on-site transverse magnetization will approach its infinite temperature expectation value ($\langle\sigma_i^z\rangle_{T=\infty}=0$) as a power law, $\frac{1}{\sqrt{\Gamma t}}$, for $\Gamma t\gg1$, when quantum coherent effects have been already exponentially suppressed by the noise. Hence the non-equilibrium dynamics of this observable is exactly the same observed for the number of kinks. In the next section we are going to consider two-points functions of the transverse magnetization looking for new physics behind the interplay of noise and quench.

\begin{figure}[htbp]
\centering
\includegraphics[scale=0.8]{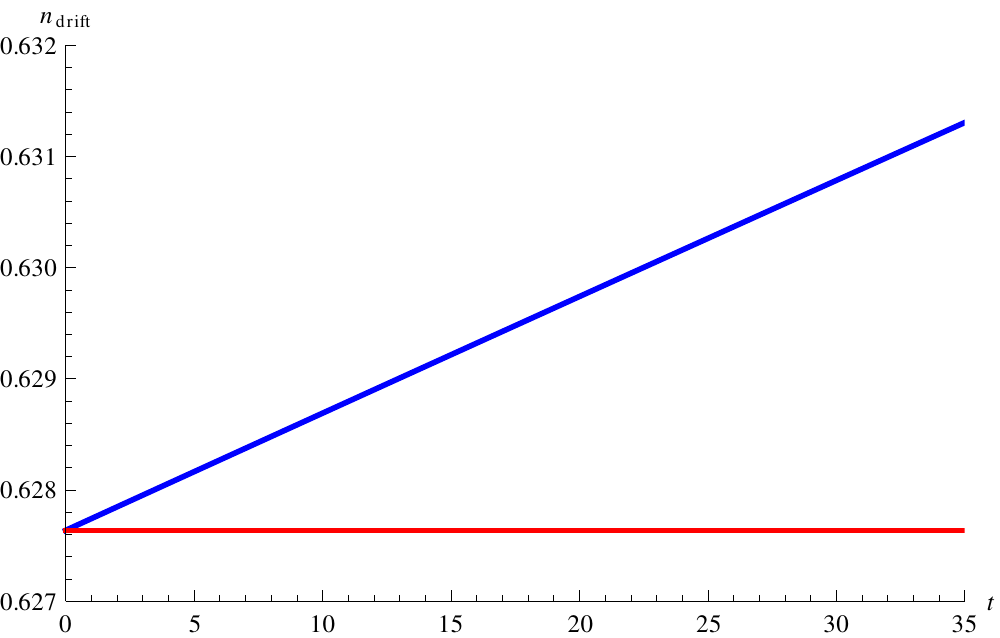}
\caption{[\it Colors online \rm]  Red line: populations contribution, $n_{drift}$, for the case of a quantum quench
($\Gamma = 0$). Blue line: populations contribution in the case of a quench with noise
($\Gamma= 0.01$). $g_0= 1.1$, $g = 4$.}
\end{figure}

\begin{widetext}

\begin{figure}[htbp]
\centering
\includegraphics[scale=0.8]{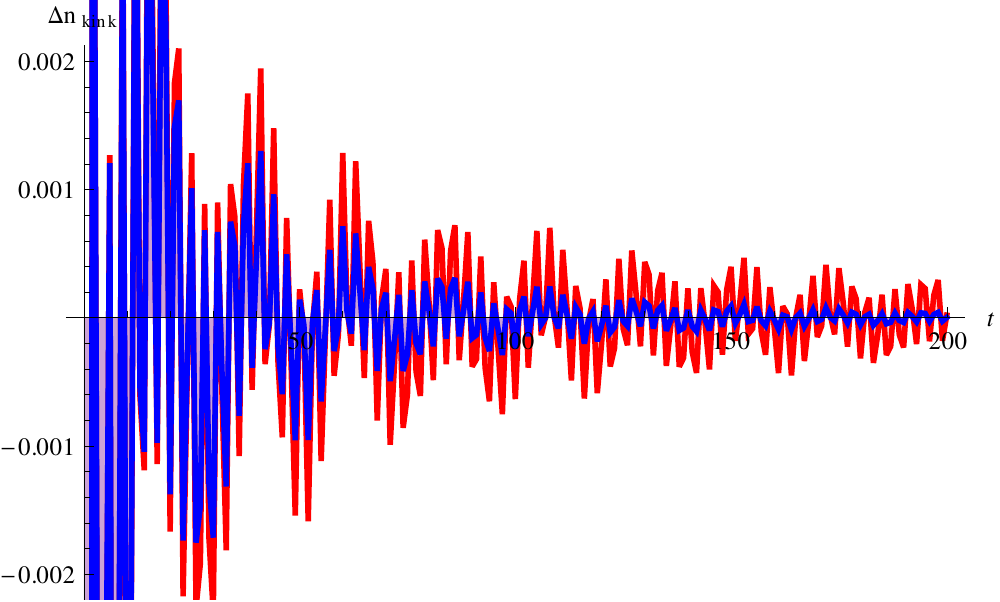}
\caption{[\it Colors online \rm]  Red line: coherences contribution, $\Delta n_{kink}$, for the case of a quantum quench
($\Gamma = 0$). Blue line: coherences contribution in the case of a quench with noise
($\Gamma= 0.01$). $g_0= 1.1$, $g = 4$.}
\end{figure}

\end{widetext}

\subsection{Transverse magnetization correlator}

A similar scenario can be also observed in the equal-time transverse magnetization correlation function, computed at different spin sites $\rho^{zz}(r,t)=\langle\sigma_{i+r}^z(t)\sigma_{i}^z(t)\rangle$.
Similarly to what we have done for $n_{kink}$, the expression for $\rho^{zz}(r,t)$ can be written as a sum of three terms
\begin{equation}\begin{split}
&\rho^{zz}(r,t)=\langle \sigma_r(t)\sigma_0(t)\rangle_{pop.}+\\&+\langle \sigma_r(t)\sigma_0(t)\rangle_{coh.}+\langle \sigma_r(t)\sigma_0(t)\rangle_{mix.}
\end{split}\end{equation}
where
\begin{equation}\label{pop}\begin{split}
&\langle \sigma_r(t)\sigma_0(t)\rangle_{pop.}=4\int_{-\pi}^{\pi}\frac{dk}{2\pi}\int_{-\pi}^{\pi}\frac{dk'}{2\pi}e^{i(k-k')r}\times\\&\times\Big[\sin{2\theta_k}\sin{2\theta_{k'}}\delta f_{k}(t)\delta f_{k'}(t)+\\
&+\Big(\frac{1}{2}+\cos{2\theta_{k'}}\delta f_{k'}(t)\Big)\Big(\frac{1}{2}-\cos{2\theta_{k}}\delta f_k(t)\Big)\Big],
\end{split}\end{equation}
\begin{equation}\label{coh}
\begin{split}
&\langle \sigma_r(t)\sigma_0(t)\rangle_{coh.}=4\int_{-\pi}^{\pi}\frac{dk}{2\pi}\int_{-\pi}^{\pi}\frac{dk'}{2\pi}e^{i(k-k')r}\times\\&\times\Big[-\sin{2\theta_k}\sin{2\theta_{k'}}y_{k}(t)y_{k'}(t)+(x_k(t)+iy_k(t)\cos{2\theta_k})\times \\&\times( x_{k'}(t)-iy_{k'}(t)\cos{2\theta_{k'}})\Big],
\end{split}
\end{equation}
\begin{equation}\label{mix}
\begin{split}
&\langle \sigma_r(t)\sigma_0(t)\rangle_{mix.}=4\int_{-\pi}^{\pi}\frac{dk}{2\pi}\int_{-\pi}^{\pi}\frac{dk'}{2\pi}e^{i(k-k')r}\times \\
&\times\Big[i\delta f_k(t)\sin{2\theta_k}(x_{k'}(t)-y_{k'}(t)\cos{2\theta_{k'}})+\\&-i\delta f_{k'}(t)\sin{2\theta_{k'}}(x_{k}(t)+ y_{k}(t)\cos{2\theta_{k}})+\\
&+\sin{2\theta_k}\delta f_{k'}(t)y_{k}(t)\cos{2\theta_{k'}}+\sin{2\theta_{k'}}\delta f_{k}(t)y_{k'}(t)\cos{2\theta_{k}}\Big].
\end{split}
\end{equation}
Looking the expression of the coherences \eqref{zetatot}, it should be clear that we can extract from the integrals in \eqref{coh} and \eqref{mix} a purely time dependent exponential decay prefactor, which allow us to neglect these terms in the $\Gamma t\gg1$ limit
\begin{equation}\label{diptemp}\begin{split}
&\langle \sigma_r(t)\sigma_0(t)\rangle_{coh.}\propto e^{-2\Gamma t}\\
&\langle \sigma_r(t)\sigma_0(t)\rangle_{mix.}\propto e^{-\Gamma t}.
\end{split}\end{equation}
In order to discriminate the separate physical associated to noise and to the ordinary quench dynamics, we start our analysis considering the case in which the QIC is driven oout of equilibrium only by the noise, $g_0=g$, and later we will consider the more involved case of the interplay between quench and noise.

\subsubsection{Noise without quench}

Let us assume to be in the long time limit $\Gamma t\gg1$, and let us restrict our attention to a protocol without quench ($g_0=g$). \\

The dynamics is dominated by modes near to $k=0,\pm\pi$ which have the slowest relaxation. We can thus at long times evaluate the correlator $\rho^{zz}$ as
\begin{equation}
\rho^{zz}\simeq\rho^{zz}_0+\rho^{zz}_{\pi}+\rho^{zz}_{-\pi}
\end{equation}
where the first contribution (which is also the only one that would survive in the scaling limit if taken from the outset) comes from modes close to $k\sim0$, the second and the third one come from modes close to $k\sim\pm\pi$. Let us then consider first $\rho^{zz}_0$.

Equation \eqref{pop} for large enough times $\Gamma t\gg1$ becomes
\begin{equation}\label{paracorr}
\begin{split}
 &\rho^{zz}_0(r,t) \simeq4\int_{-\infty}^{\infty}\frac{dk}{2\pi}\int_{-\infty}^{\infty}\frac{dk'}{2\pi}e^{i(k-k')r}\\
&\Big(\frac{1}{4}+\frac{k}{E_k}\delta f_k(t)\frac{k'}{E_k'}\delta f_{k'}(t)-\frac{\Delta^2}{E_k E_{k'}}\delta f_k(t)\delta f_{k'}(t)\Big),
\end{split}
\end{equation}
\\

where the time dependence of $\rho^{zz}_0(r,t)$ is going to be fully determined by the slowest mode $k\simeq0$, and where the small $k$ behaviour of $\delta f_k$ is taken
\begin{equation}\label{senzaq}
\delta f_k(t)\substack{\ \\[1mm]=\\k\simeq0}-\frac{1}{2}e^{-\Gamma t\frac{k^2}{\Delta^2}}.
 \end{equation}
The correlator can thus be derived by computing the following integral
\begin{equation}\label{fondamentale}
 I=\int_{-\infty}^{\infty}\frac{dk}{2\pi}\frac{e^{ikr}}{E_k}e^{-\Gamma t\frac{k^2}{\Delta^2}}.
\end{equation}
First of all, we make the substitution $k=\Delta q$
\begin{equation}\label{integraleq}
 I=\int_{-\infty}^{\infty}\frac{dq}{2\pi}\frac{e^{iq\Delta r}}{\sqrt{q^2+1}}e^{-\Gamma tq^2}.
\end{equation}
From Eq. \eqref{integraleq} it is clear that the exponential decay induced by the noise gives a natural cut-off which enforces the convergence of the integral; in particular, it is clear that the largest contribution to the integral comes from the modes $q\ll\frac{1}{\sqrt{\Gamma t}}$; in other words, recalling that $\Gamma t\gg1$, we can expand the denominator of the integrand for small $q$. To first order we get
\begin{equation}\label{integrale}\begin{split}
 I=&\int_{-\infty}^{\infty}dq e^{iq\Delta r}e^{-\Gamma tq^2}(1-\frac{1}{2}q^2+...)=\\
&=\sqrt{\frac{\pi}{\Gamma t}}e^{-\frac{(\Delta r)^2}{4\Gamma t}}+O\Big(\frac{\Delta r}{\Gamma t}\Big)
\end{split}\end{equation}
and so, substituing in \eqref{paracorr}, for the transverse magnetization correlator we get
\begin{equation}\label{corto}
\rho^{zz}_0(r,t)=-\frac{1}{\pi}\frac{\Delta^2}{4}\frac{1}{\Gamma t}e^{-\frac{(\Delta r)^2}{2\Gamma t}}
\end{equation}

Concerning the computation in the $\Delta r\gg\Gamma t$ regime, we observe first of all that

\begin{equation}\label{gammaf}
\frac{1}{(q^2+1)^{1/2}}=\frac{1}{\Gamma(\frac{1}{2})}\int_0^\infty daa^{-1/2}e^{-a(q^2+1)}
\end{equation}
where $\Gamma(\frac{1}{2})$ is the Euler Gamma function. Inserting \eqref{gammaf} in \eqref{integraleq}, we have
\begin{widetext}
\begin{equation}\label{bessel}\begin{split}
&\int_{-\infty}^{\infty}dq\frac{e^{iq\Delta r-\Gamma tq^2}}{\sqrt{q^2+1}}=\int_{-\infty}^{\infty}dq\int_0^\infty\frac{da}{\Gamma(1/2)}a^{-1/2}e^{-iqmr-\Gamma tq^2-a(q^2+1)}
=\int_0^\infty daa^{-1/2}e^{-\frac{m^2r^2}{4(a+\Gamma t)}-a}\frac{1}{\sqrt{a+\Gamma t}}=\\
&\substack{\ \\[1mm]=\\a+\Gamma t\equiv b}\int_{\Gamma t}^\infty\frac{db}{\sqrt{b-\Gamma t}}\frac{e^{-\frac{m^2r^2}{4b}-b+\Gamma t}}{\sqrt{b}}\substack{\ \\[3mm]=\\b\equiv\frac{mr}{2}c}e^{\Gamma t}\int_{2\frac{\Gamma t}{mr}}^\infty dc\frac{\sqrt{\frac{mr}{2}}}{\sqrt{\frac{mrc}{2}-\Gamma t}}\frac{1}{\sqrt{c}}e^{-mr(c+\frac{1}{c})}=2e^{\alpha^2\beta}\int_{\alpha}^\infty dx\frac{1}{\sqrt{x^2-\alpha^2}}e^{-\beta \Big(x^2+\frac{1}{x^2}\Big)}
\end{split}\end{equation}
\end{widetext}

where in the last equality we defined $c=x^2$, $\alpha^2=\frac{2\Gamma t}{\Delta r}$ and $\beta=\frac{\Delta r}{2}$. The last integral in Eq. \eqref{bessel} can be evaluated with a saddle point approximation around $x\simeq1$, in the limit $\alpha\ll1$, $\beta\gg1$
\begin{equation}\begin{split}
&2e^{\alpha^2\beta}\int_{\alpha}^\infty dx\frac{1}{\sqrt{x^2-\alpha^2}}e^{-\beta \Big(x^2+\frac{1}{x^2}\Big)}\simeq\\
&\simeq2e^{\alpha^2\beta}\frac{e^{-2\beta}}{\sqrt{1-\alpha^2}}\int_0^\infty dxe^{-4\beta(x-1)^2}\substack{\ \\[2mm]\simeq\\ \beta\gg1}\\
&\simeq\frac{2e^{\alpha^2\beta-2\beta}}{\sqrt{1-\alpha^2}}\frac{\sqrt{\pi}}{2\sqrt{\beta}}=\sqrt{\frac{2\pi}{\Delta r}}\frac{e^{-\Delta r+\Gamma t}}{\sqrt{1-\frac{2\Gamma t}{\Delta r}}}\substack{\ \\[2mm]\propto\\ \frac{\Delta r}{\Gamma t}\gg1}\frac{e^{-\Delta r}}{\sqrt{\Delta r}}.
\end{split}\end{equation}
where we kept the gaussian fluctuations around the saddle point $x\simeq1$.\\

This expression allows to find the correlation function in the $\Delta r\gg\Gamma t$ limit, after some straightforward algebra on Eq.\eqref{paracorr}
\begin{equation}\label{grandi}
\rho^{zz}_0(r,t)\simeq \frac{e^{-2\Delta r}}{2\pi r^2}.
\end{equation}

It should be clear from these expressions that the diffusive behaviour found for the correlator \eqref{corto} in the $\Delta r\ll \Gamma t$ limit and indicating the continous heating of the system towards the infinite temperature state, travels with a wavefront speed $\gamma=\frac{\Gamma}{\Delta}$, which means that points with $\Delta r\gg\Gamma t$ do not present any signature of the noise and their correlation function is the same of $\sigma_i^z$ in the QIC without noise and quench (see eq. \eqref{grandi} and for comparison \cite{Sachdev1999}).

Before considering the combined signature of the noise and the quench on the on-site magnetization correlation function, let us restore  lattice corrections originating from $k\simeq\pm\pi$ modes in Eq. \eqref{pop}; for $\rho^{zz}(r,t)$, in the $\frac{\Delta r}{\Gamma t}\ll1$ limit, we get (assuming the lattice spacing $a=1$)
\begin{equation}
\rho^{zz}(r,t)=-\frac{1}{\pi}\frac{\Delta^2}{4}\frac{1}{\Gamma t}e^{-\frac{(\Delta r)^2}{2\Gamma t}}\Big(1+\frac{g+1}{g-1}\cos(\pi r)e^{-\frac{gr^2}{\Gamma t}}\Big)^2.
\end{equation}
In the space-time region defined by $\sqrt{\frac{\Gamma t}{g}}\ll r\ll \frac{\Gamma t}{\Delta}$, lattice corrections are completely negligible, on the other hand, in the limit $r\ll\sqrt{\frac{\Gamma t}{g}}$ the signature of the noise is still diffusive. 
Therefore, we can conclude that the qualitative behaviour of the on-site magnetization correlation function is diffusive.

\subsubsection{Effect of the quench}

Now we are interested in studying the interplay between quench and noise in the spreading of quantum correlations in $\rho^{zz}(r,t)$.
We use the expressions for populations and coherences, \eqref{esatta}, \eqref{zetatot}, and look for the different spatio-temporal regimes emerging during the time evolution of this observable.

The dynamics is characterized by the propagation of two ``wave'' fronts: at earlier times, $\Gamma t\ll1$, a first front appears at $r\simeq Jt$, controlled by the velocity of quasiparticles emitted after a quench ($v\simeq J$), which separates unconnected space-time regions, $r\gg Jt$, where $\sigma_i^z$ correlations behave as in the QIC without quench, from a region of space-time connected points $r\ll t$, where the stationary correlation function is the same of a quenched QIC \cite{EsslerCalabreseFagotti}. This is consistent with the Lieb-Robinson limit \cite{LB}, as already found for other systems \cite{Polkovnikov2010} and by many authors for the sudden quench of the QIC \cite{Rossini, EsslerCalabreseFagotti,Ising}. The effects of the noise are hardly relevant at early times as observed for the evolution of $n_{kink}$.

On the other side, taking the long time limit, $\Gamma t\gg1$, for $\Delta r\ll\Gamma t$ we find again a diffusive spreading of correlations, while for unconnected spacetime points ($\Delta r\gg\Gamma t$) the stationary correlation function crosses over to the asymptotic expression of the correlation function in a quenched QIC without noise \cite{EsslerCalabreseFagotti}.

This scenario can be summarized in the following expressions for the correlation function

\begin{equation}
\rho^{zz}(r,t)\simeq_{\Gamma t\ll1}\begin{cases}
\frac{1}{2\pi r^2}\exp[-2\Delta_0r] & \text{$r\gg vt$}\\
& \\
\frac{1}{r^\alpha}\exp[-r/\xi_z] & \text{$r\ll vt$}
\end{cases}
\end{equation}
where $\xi_z$ is the correlation lenght associated to a simple \emph{quantum quench} of the transverse field and $\alpha$ a constant, computed in \cite{EsslerCalabreseFagotti}.
In the large-times regime, $\Gamma t\gg1$, the noise becomes relevant and the second crossover, between quenched QIC correlation functions and \it diffusive \rm behavior emerges
\begin{equation}
\label{cortoq}
\rho^{zz}(r,t)\simeq_{\Gamma t\gg1}\begin{cases}
\frac{1}{r^\alpha}\exp[-r/\xi_z] & \text{$\gamma t\ll r \ll vt$}\\
& \\
-\frac{1}{\pi}\frac{\Delta^2}{4}\frac{1}{\Gamma t}\exp\left[-\frac{(\Delta r)^2}{2\Gamma t}\right] & \text{ $r\ll \gamma t$}
\end{cases}
\end{equation}
where $\gamma=\frac{\Gamma}{\Delta}$ is the small parameter, which controls the self-consistent Born approximation used in Section IV to resum the Dyson series \cite{note1}.
\begin{center}
\begin{figure}[htbp]
\includegraphics[scale=0.37]{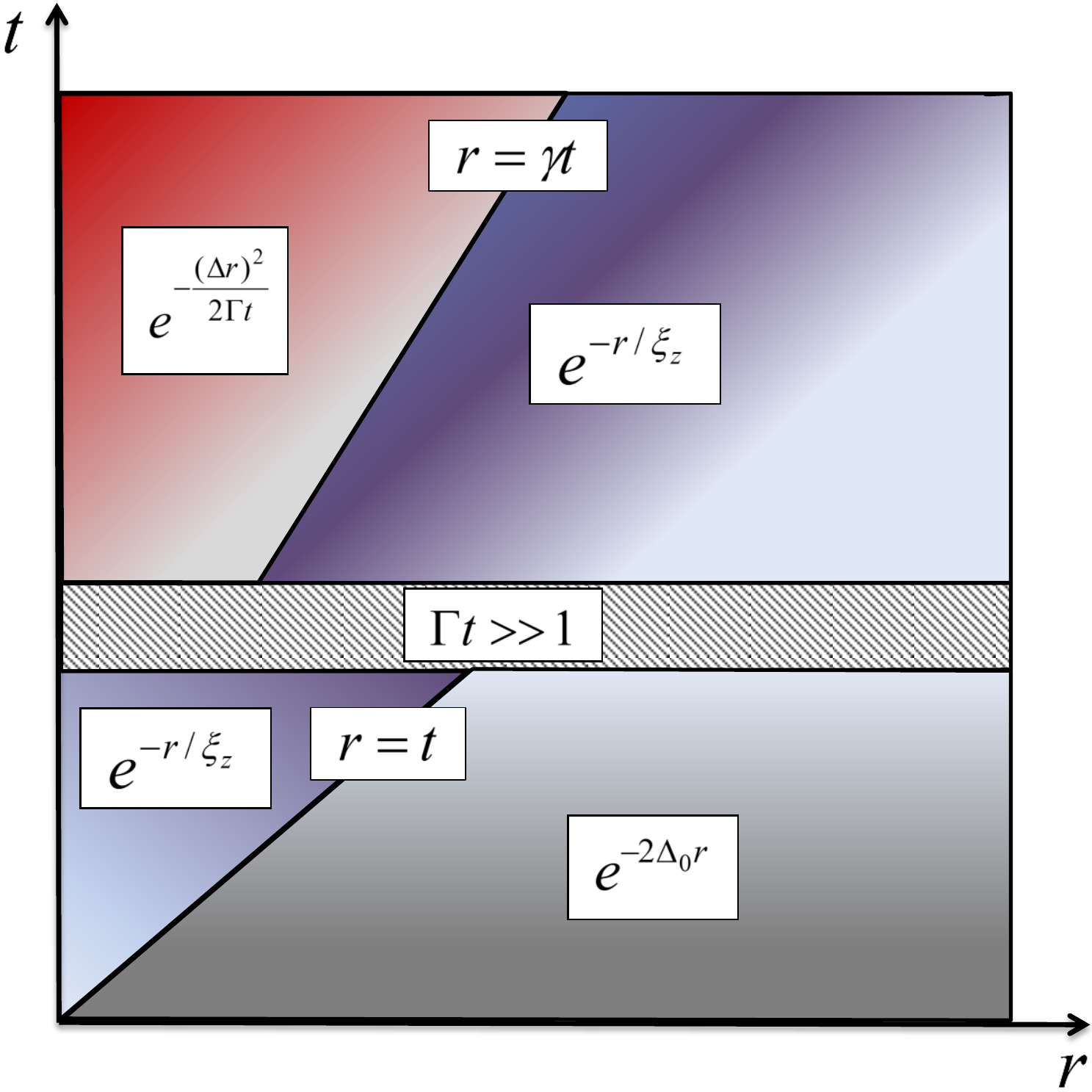}
\caption{The spreading of quantum and thermal correlations in the noisy Quantum Ising Model ($J=1$): the transverse field correlator has a first crossover when ballistic quasi-particles, carrying quantum correlations, propagate at the distance $r$. Thermal correlations propagate at a second stage, leading to a crossover to a diffusive form, consistent with thermal dynamics. }\label{cono}
\end{figure}
\end{center}

This type of ligt-cone spreading of correlations has been observed experimentally (without noise) in the quench dynamics of the Bose-Hubbard model \cite{Kollath} and in the coherent split of 1D Bose gases, characterizing the wave front associated to the pre-thermal state \cite{Geiger}. \\

\subsection{Order Parameter correlations}

This last subsection is devoted to study whether the diffusive behaviour observed before is a general signature of the effect of the noise in correlation functions; in order to answer to this question, it is sufficient to compute the equal-time order parameter correlation functions, $\rho^{xx}_{lm}$ for a QIC perturbed by the noise without adding the effect of a quench in the transverse field.

The usual way to perform this computation in the ground as in a thermal state is to recast $\rho^{xx}_{lm}$ in a Toepltiz determinant form and to evaluate the large-spin separation limit $l-m=n\rightarrow\infty$, using Fisher-Hartwig conjecture \cite{FH}. For a quantum quench the situation is in general much more complicated \cite{EsslerCalabreseFagotti}. Hence, we will therefore restrict our attention to the dynamics in the presence of the noise at long-times where the coherences have been suppressed and only populations evolve. In this case we may proceed with standard methods.

Let introduce the operators
\begin{equation}\label{JW}
A_i\equiv c_i^{\dag}+c_i\quad B_i\equiv c_i^{\dag}-c_i
\end{equation}
where $c_i$ is the Jordan-Wigner fermion on the lattice; from \eqref{JW} it follows \cite{Toep} that
\begin{equation}\label{ordine}\begin{split}
\rho^{x}_{ml}&=\langle\sigma_m^x\sigma_l^x\rangle=\\
&=\langle B_lA_{l+1}B_{l+1}...A_{m-1}B_{m-1}A_{m}\rangle
\end{split}
\end{equation}
We can factorize this expression, using Wick theorem, and, noticing that $\langle A_l A_m\rangle=0$ and $\langle B_lB_m\rangle=0$, we only need to compute $\langle B_lA_m\rangle$:
\begin{equation} \label{mon}
\langle B_lA_m\rangle=\int_{-\pi}^{\pi}\frac{dk}{2\pi}e^{-ikR}e^{i2\theta_k}2\delta f_k\equiv s(R)
\end{equation}
where $R=l-m$ and $\delta f_k=-\frac{1}{2}e^{-\Gamma\sin^22\theta_1t}$.

It is possible to show \cite{Toep} that the order parameter correlator can be cast in the form of a $n+1 \times n+1$ \emph{Toeplitz determinant}
\begin{equation}\label{determinante}
det(T_n)=det\arrowvert s(j-k)\arrowvert_{j,k=0}^{n}=D_n[f]
\end{equation}
where $T_n$ is a Toeplitz matrix
\begin{equation}
T_n=\begin{pmatrix}
     s(0) & s(-1) & s(-2) & ...& s(-n)\\
     s(1) & s(0) & s(-1) & ...& s(1-n)\\
     s(2) & s(1) & s(0) & ...& s(2-n)\\
     ... & ... & ... & ... & ...\\
     s(n) & s(n-1) & s(n-2) & ...& s(0)
    \end{pmatrix}
\end{equation}
It is convenient to write
\begin{equation}
s(R)=\int_{-\pi}^{\pi}\frac{dk}{2\pi}e^{-iRk}f(k)
\end{equation}
where $f(k)$ is a periodic complex function $f(k)=f(k+2\pi)$, called the \emph{generating function}. \\

Let us now compute the order parameter correlator (Eq. \eqref{ordine}) in the large $R$ limit, using the large $n$ expansion of a Toeplitz determinant (Eq. \eqref{determinante}) which can be exctracted using the Fisher-Hartwig conjecture \cite{FH}. The latter states that, if $f(k)$ can be cast in the form

\begin{widetext}
\begin{equation}\label{decomp}
f(k)=f_0(k)\prod_r \exp{[ib_r(k-k_r-\pi sign(k-k_r))]}(2-2\cos{(k-k_r)})^{a_r}
\end{equation}
\end{widetext}

where $k\in(0,2\pi)$, $k_r$ are singularities (jumps, zeros or poles) of $f(k)$, $f_0(k)$ is an infinitely differentiable function in $k\in(0,2\pi)$ and $a_r,b_r$ are two complex numbers, then the asymptotic expansion of the Toeplitz determinant, for large $n$, is
\begin{equation}\label{asint}
T_n\substack{\ \\[2mm]\sim\\ n\rightarrow\infty}e^{l_0n}n^{\sum_r(a_r^2-b_r^2)},
\end{equation}
where $l_0=\int_{-\pi}^{\pi}\frac{dk}{2\pi}\log{f_0(k)}$. \\

First of all, we are briefly going to set the notation, computing the order parameter correlator of the QIC at equilibrium, and then we will move to the case of interest for this Section.

\subsubsection{Order Parameter Correlations in the QIC}
Consider the Quantum Ising Model
\begin{equation}
H_0=-\sum_i\sigma_i^x\sigma_{i+1}^x+g\sigma_i^z
\end{equation}
in the paramagnetic phase $g>1$.\\

In this case (see note \cite{nota5})
\begin{equation}\label{eq}\begin{split}
\langle B_l& A_m\rangle=s(R)=\\
=&\int_{-\pi}^{\pi}\frac{dk}{2\pi}e^{ikR}e^{-ik}\frac{\cos{k}-g+i\sin{k}}{\sqrt{(\cos{k}-g)^2+\sin^2{k}}}
\end{split}
\end{equation}
and $f(k)$ can be rewritten, making the change of variable $z=e^{ik}$, as a function in the complex plane ($\lambda\equiv1/g$)
\begin{equation}\label{comp}
f(z)=z^{-1/2}\frac{(z-g)^{1/2}}{(zg-1)^{1/2}}=z^{-1}\frac{(\lambda z-1)^{1/2}}{(\lambda z^{-1}-1)^{1/2}},
\end{equation}
which has four branch points $z=0$, $1/g$, $g$, $\infty$. We choose the two branch cuts in the following way: the first linking $z=0$ with $z=1/g$, and the second one linknig $z=g$ with $z=\infty$. \\

It is not immediate to apply the Fisher-Hartwig conjecture on Eq.\eqref{comp}; in this case, some additional manipulations on the generating function are required, following Ref. \cite{Frankel}, we note that
\begin{equation}\label{contorno}
\frac{1}{2\pi}\int_{-\pi}^{\pi}dk f(k) e^{-ikR}=\int_{\emph{C}}f(z)z^{-R}\frac{dz}{2\pi iz}
\end{equation}
where $\emph{C}$ is a closed contour encircling the origin in the anulus $1/g<|z|<g$, where $f(z)$ is analytic with our choices of branch cuts. The integral involed in the Toeplitz Determinant is defined over a circle of radius 1, encircling the origin, \eqref{contorno}, but applying Cauhy's theorem inside the anulus $1/g<|z|<g$ we can move the integration from the circle of radius 1 to the circle of radius $g=1/\lambda$; this is equivalent to make the substituion $z\rightarrow z/\lambda$ in \eqref{contorno}, and to keep the integration over the circle $\arrowvert z\arrowvert=1$, as shown in \cite{Frankel} (for a technical remark on this point see \cite{nota6}).\\

Following this procedure it is possible to rewrite Eq. \eqref{comp} in this form
\begin{equation}
f(z)=\frac{\lambda}{z}\frac{(1-z)^{1/2}}{\Big(1-\frac{\lambda^2}{z}\Big)^{1/2}},
\end{equation}
where the Fisher-Hartwig formula can be immediately applied; resubstituting again $z=e^{ik}$, we get the following Fisher-Hartwig decomposition \eqref{decomp}
 \begin{equation}
f(k)\sim f_0(k)e^{-\frac{3}{4}ik}(1-\cos{k})^{1/4},
 \end{equation}
where
\begin{equation}
f_0(k)=\frac{\lambda}{(1-\lambda^2e^{-ik})^{1/2}}.
\end{equation}
It is now easy to show that
\begin{equation}
l_0=\int_{-\pi}^{\pi}\frac{dk}{2\pi}\log{\frac{\lambda}{(1-\lambda^2e^{-ik})^{1/2}}}=\log\lambda,
\end{equation}
which gives for the correlation function $\rho^{xx}(R)$, according to \eqref{asint}, the following result
\begin{equation}\label{xxeq}
\rho^{xx}(R)\substack{\ \\[1mm]\sim\\ R\rightarrow\infty} R^{-1/2} e^{-R/\xi_{eq}}
\end{equation}
where $\xi_{eq}=(\log g)^{-1}$.

\subsubsection{Order Parameter correlations in a noisy QIC}
We are now ready to derive the main result of this section, adding to the QIC the usual noisy time dependent perturbation.
Recalling \eqref{mon}, we get in this case for the generating function $f(k)$
\begin{equation}\label{generatore}
f(k)=e^{-\Gamma t\sin^22\theta_k}f^{eq}(k),
\end{equation}
where $f^{eq}(k)$ is the static generating function for the Toepltiz determinant in the QIC at equilibrium, presented in the previous subsection. The function $e^{-\Gamma t\sin^22\theta_k}$ is non zero and smooth in $(0,2\pi)$, so our only task is to make the change of variable in the complex plane $z\rightarrow z/\lambda$ as before, necessary to apply the Fisher Hartwig conjecture.

The correlation function, using Fisher-Hartwig conjecture, takes the form
\begin{equation}
\rho^{xx}(R,t)\substack{\ \\[1mm]\sim\\ R\rightarrow\infty}R^{-1/2} e^{-R/\xi(t)},
\end{equation}
where
\begin{equation}
\frac{1}{\xi(t)}=\frac{1}{\xi_{eq.}}+\frac{1}{\xi(t)_{noise}}
\end{equation}
and
\begin{equation}
\frac{1}{\xi(t)_{noise}}=\Gamma t\int_{-\pi}^{\pi}\frac{dk}{2\pi}a(k);
\end{equation}
$\xi_{eq.}$ is the exponent coming from the regular part of the generating function at equilibrium (see Eq. \eqref{xxeq}), while $a(k)$ has the following form
\begin{equation}
a(k)\equiv\sin^22\theta_k=\frac{(e^{ik}-e^{-ik})^2}{(e^{ik}-e^{-ik})^2-(2g-e^{ik}-e^{-ik})^2}.
\end{equation}

The integral
\begin{equation}
\int_{-\pi}^{\pi}\frac{dk}{2\pi}a(k)
\end{equation}
can be written in the complex plane ($z=e^{ik}$) as
\begin{equation}\label{complesso}
\oint_{|z|=1}\frac{dz}{2\pi iz} a(z),
\end{equation}
where
\begin{equation}
a(z)\equiv\frac{1}{1-\Big(\frac{(z-1-\sqrt{1-\frac{1}{g^2}})(z-1+\sqrt{1+\frac{1}{g^2}})}{(z-\frac{1}{g})(z+\frac{1}{g})}\Big)^2}
\end{equation}
has poles in $z=0,\frac{1}{g^2}, 1$.

Considering we move from the circle of radius 1 to the one of radius $\frac{1}{\lambda}$, where $f_{eq}(k)$ has a branch cut, we need to regularize the integral \eqref{complesso}, deforming the integration countour from inside in order to avoid $z=1$; in other words, we consider the circle of radius $1-\epsilon$, taking the limit $\epsilon\rightarrow0^+$.

Applying the residue theorem to \eqref{complesso} we get
\begin{equation}
\frac{1}{\xi(t)_{noise}}=\frac{\Gamma t}{2g^2}
\end{equation}

This result can be checked numerically, studying the asymptotic behaviour of a Toeplitz determinant, whose entries are generated by \eqref{generatore}.

For a quench without dissipation the stationary correlation function has in general an exponential form
$\rho^{xx}(R,t)\sim \exp[- R/\xi]$,
with a correlation length $\xi$ dictated by the non-thermal distribution function of quasi-particles and predicted by the Generalized Gibbs ensemble \cite{EsslerCalabreseFagotti}. Turning on the noise, the signatures of the crossover observed for the transverse magnetization are expected in this case to be different; indeed, the same exponential form persists and the spreading of quantum and thermal correlations will not result in a diffusive form, but rather modify just the specifics of the correlation length which at later times shrinks as $1/\Gamma t$ for large times.

The different signatures observed in the transverse and longitudinal magnetization are consistent with analogous phenomenology observed elsewhere for quenches in the QIC~\cite{Rossini}.

\section{Conclusions}

In this paper we studied the effect of the noise on the non-equilibrium dynamics of a Quantum Ising Chain driven out of equilibrium by a sudden quench of the transverse field. We considered a gaussian time-dependent delta correlated noise superimposed on top of the transverse magnetization, generalizing in this way to the noisy case the ordinary sudden quench dynamics addressed in other works \cite{Rossini,EsslerCalabreseFagotti,Ising}. First of all, we computed in the small noise limit $\frac{\Gamma}{\Delta}\ll1$, the statistic of the work done on the system for static and dynamical noisy out-of-equilibrium protocols, showing in the static case that the effect of the fluctuations is to smooth the singularities associated to the existence of a low-energy quasiparticle production threshold in the usual sudden quench of the QIC, while in the dynamical case we have shown the additional emergence of a time-dependent spectral weight of the edge singularity in $P(\omega,\tau)$. \\

The non-equilibrium dynamics resulting from the interplay of a quantum quench and a time dependent noise is characterized by three stages. First, inhomoegenous dephasing brings the system towards the GGE of the unperturbed Ising chain; then, a second dephasing mechanism comes into play, killing exponentially the coherences on the time scale of the inverse noise amplitude. Finally, the noise heats up the populations, driving the system towards the infinte temperature state, as confirmed by the study of a wide class of observables (number of kinks, on-site transverse magnetization, correlation function of the transverse magnetization). It is a remarkable fact that analogously to non-integrable quantum many body systems \cite{Kollar2011}, an intermediate steady state appears during dynamics, which can be considered in a broader sense a \emph{prethermal} state. We found, remarkably, that this generalized \emph{prethermalization} occurs only in those observables which can show an interplay between the relaxation and dephasing of populations and coherences.\\

We conclude observing that the method, used in this paper and based on Keldysh tecnique, could be employed also to understand thermalization dynamics of quenched closed quantum many body systems \cite{Marcuzzi}, where many issues concerning prethermalization and the role of interactions in out-of-equilibrium problems, such as the time scales involved, are still elusive; finally, the noisy Quantum Ising Chain is a potential playground to study fluctuation-dissipation relations out-of-equilibrium, which has been recently at the centre of the attention of a series of papers on this topic \cite{Ising}.\\

\section{Acknowledgements}

We would like to thank G. Biroli, P. Calabrese, M. Fabrizio, R. Fazio, G. Mussardo, G. Santoro for comments and in particular A. Gambassi and P. Smacchia for helpfull discussions. JM is indebted with M. Marcuzzi for helpfull comments on Eq.\eqref{bessel}.

We would like to thank the KITP in Santa Barbara for hospitality during the workshop on ”Quantum Dynamics in Far from Equilibrium Thermally Isolated Systems”. This research was supported in part by
the National Science Foundation under Grant No. NSF
PHY11-25915.

\widetext

\begin{appendix}

\section{Generalized time-dependent Bogolyubov transformation and statistics of the work in the Quantum Ising Chain}
In this Appendix we derive a formula for the characteristic function, $G(u)$, introduced in Section III, Eq. \eqref{char}. We use a generalization of Bogolyubov transformations for time-dependent protocols, and then in Section III we specialize these results for a time-dependent noisy perturbation.

We consider a QIC in the transverse field $g_0$ and we prepare the system in the ground state of the paramagnetic phase, $|\psi(g_0)\rangle$; we perform a generic time-dependent protocol, $g(t)$, with these boundary conditions: $g(t=0)=g_i>1$ and $g(t=\tau)=g_f>1$, in general $g_i, g_f\neq g_0$. For instance, the sudden quench case is recovered from our expressions when $\dot{g}(t)=0$, hence $g_i=g_f$.

Our goal is to compute
\begin{equation}\label{carfun}
 G(u)=\langle\psi(g_0)|e^{iuH^H_{\tau,\tau_0}}|\psi(g_0)\rangle,
\end{equation}
where $H^H_{\tau,\tau_0}=U^\dag(\tau,\tau_0)H_{\tau,\tau_0}U(\tau,\tau_0)$ denotes the Hamiltonian used in the measurement process; the superscript $H$ indicates that operators are taken in the Heisenberg picture. In Eq. \eqref{carfun} we dropped the inessential global phase prefactor present in Eq. \eqref{char}. We can rewrite $G(u)$ in Schrodinger representation, absorbing the evolution in the wavefunction $|\psi(\tau)\rangle=U(\tau,\tau_0)|\psi(g_0)\rangle$,
\begin{equation}
 G(u)=\langle\psi(\tau)|e^{iuH_{\tau}}|\psi(\tau)\rangle.
\end{equation}
In order to compute this quantity, we make the central \emph{ansatz} of our method, that consists in introducing an operator
$\widetilde{\gamma_k}(t)$, which annhilates the state at time $t$
\begin{equation}
\widetilde{\gamma_k}(t)|\psi(t)\rangle=0,
\end{equation}

which means that $|\psi(t)\rangle$ is a \emph{Bogolyubov vacuum} at each time, for a certain operator, $\widetilde{\gamma_k}(t)$. The choice of the intial state implies $\widetilde{\gamma_k}(0)=\gamma_k(g_0)$.
From our ansatz, it follows that
\begin{equation}
0=i\frac{d}{dt}(\widetilde{\gamma_k}(t)|\psi(t)\rangle)=\Big(i\frac{\partial}{\partial t}\widetilde{\gamma_k}(t)
\Big)|\psi(t)\rangle+\widetilde{\gamma_k}(t)\Big(i\frac{\partial}{\partial t}|\psi(t)\rangle\Big)=\Big(i\frac{\partial}{\partial t}\widetilde{\gamma_k}(t)+\widetilde{\gamma_k}(t)H(t)-H(t)\widetilde{\gamma_k}(t)\Big)|\psi(t)\rangle
\end{equation}
and this implies
\begin{equation}\label{evoluzione}
i\frac{\partial}{\partial t}\widetilde{\gamma_k}(t)=-[\widetilde{\gamma_k}(t),H(t)].
\end{equation}
At a certain time $t$, $H(t)$ is diagonalized by a set of Bogolyubov operators $\gamma_k(t)$, which are related in the usual way to the Jordan-Wigner fermions, $c_k=u_k(t)\gamma_k(t)-iv_k(t)\gamma_{-k}^\dag(t)$, where $u_k(t)=\cos\theta_k(t)$, $v_k(t)=\sin\theta_k(t)$; the Bogolyubov angle, $\theta_k(t)$, depends on the time protocol $g(t)$ and the Hamiltonian is diagonalized as usual,
\begin{equation}\label{hamt}
 H(t)=\sum_{k>0}E_k(t)(\gamma_k^\dag(t)\gamma_k(t)-\gamma_{-k}(t)\gamma_{-k}^\dag(t)).
\end{equation}
Now, we looks for two time-dependent coefficients $a_k(t)$ and $b_k(t)$, which unitarly relate $\widetilde{\gamma_k}(t)$ to $\gamma_k(t)$, through the following rotation
\begin{equation}\label{bogo}
\widetilde{\gamma_k}(t)=a_k(t)\gamma_k(t)-ib_k(t)^*\gamma_{-k}^\dag(t).
\end{equation}
At $t=0$ this equation becomes with our boundary conditions
\begin{equation}
\widetilde{\gamma}_k(g_0)=a_k(t=0)\gamma_k(g_i)-ib_k(t=0)^*\gamma_{-k}(g_i)^\dag,
\end{equation}
which is the usual Bogolyubov rotation in the case of a sudden quench in the QIC (see, for instance, \cite{Silva}), with initial conditions, $a_k(t=0)=\cos\Delta\theta_k$ and $b_k(t=0)=\sin\Delta\theta_k$.
We are now ready to substitue \eqref{hamt} and \eqref{bogo} in \eqref{evoluzione}, where we need $\dot{u}_k=-v_k(t)\dot{\theta}_k(t)$ and $\dot{v}_k=u_k(t)\dot{\theta}_k(t)$; after staightforward algebra we get two coupled first order differential equations for $a_k(t)$ and $b_k(t)$
\begin{equation}\begin{split}
i\dot{a_k}&=-E_k(t)a_k-ib_k^*\dot{\theta}_k(t)\\
i\dot{b_k^*}&=i\dot{\theta}_k(t)a_k+b_k^*E_k(t).
\end{split}\end{equation}
Defining $q_k(t)\equiv\frac{b_k^*(t)}{a_k(t)}$, it is possible to write the following differential equation
\begin{equation}\label{eqdiff}
i\dot{q}_k=i\dot{\theta}_k(t)+2q_kE_k(t)+iq_k^2\dot{\theta}_k(t).
\end{equation}
In the following we will need the small $k$ expansion of $q_k$, so we solve \eqref{eqdiff}, expanding $q_k$ in series, $q_k(t)=\sum_{n=0}^\infty c_n(t)k^n$. \\

The zeroth order solution is null
\begin{equation}\begin{split}
&i\dot{c}_0=2c_0\Delta(t)\\
&c_0(t=0)=0,\\
\end{split}\end{equation}
because $q_k(t=0)=\tan\Delta\theta_k\sim_{k\sim0}\frac{1}{2}k\frac{\Delta_0-\Delta_i}{\Delta_0\Delta_i}$ has a vanishing zero order in the $k$ expansion. \\
For the first order solution we have
\begin{equation}\label{eqdiffuno}\begin{split}
i\dot{c}_1(t)&=-\frac{i}{2\Delta(t)^2}\dot{g}(t)+2\Delta(t)c_1(t)\\
c_1(t=0)&=\frac{1}{2}k\frac{\Delta_0-\Delta_i}{\Delta_0\Delta_i}.
\end{split}                                                                                                                                      \end{equation}
Using the method of separation of arbitrary constants and taking into account that $c_0(t)=0$, $\forall t$, we find
\begin{equation}\label{soluzione}
c_1(t)=e^{-2i\int_0^t\Delta(t')dt'}\Big(c_1(0)-\int_0^t\frac{e^{2i\int_0^{t'}\Delta(t'')dt''}}{2\Delta(t')^2}\dot{\Delta}(t')dt'\Big)
\end{equation}

If we now come back to the original problem, we see that \eqref{bogo} and the ansatz $\widetilde{\gamma_k}(t)|\psi(t)\rangle=0$ allows us to write the state at time $t=\tau$ as a BCS-state, similarly to what is usually done for a sudden quench in the QIC (see for instance \cite{Rossini, EsslerCalabreseFagotti, Silva} or Eq. \eqref{iniziale}):
\begin{equation}
|\psi(\tau)\rangle=\exp\Big[i\sum_{k>0}\frac{b_k(\tau)^*}{a_k(\tau)}\gamma_k^\dag(\tau)\gamma_{-k}^\dag(\tau)\Big]|0\rangle_{\tau},
\end{equation}
where $|0\rangle_{\tau}$ is the vacuum of the QIC at time $\tau$ and $\gamma_k^\dag(\tau)$, the Bogolyubov operators diagonalizing the Hamiltonian at time $t=\tau$. Following the same procedure of \cite{Silva}, it is possible to write the characteristic function, $G(u)$, of the statistics of the work as
\begin{equation}\label{caratteristicalavoro}
G(u)\sim\frac{\exp{\Big(N\int_0^\pi\frac{dp}{\pi}\log(1+|q_p(\tau)|^2e^{2iuE_p(g_f)})\Big)}}{\exp{\Big(N\int_0^\pi\frac{dp}{\pi}\log(1+|q_p(\tau)|^2)\Big)}}
\end{equation}
where $g_f=g(t=\tau)$. \\

Considering that $G(u)$ is the Fourier transform of $P(\omega)$, and since we are interested in the low energy behaviour of $P(\omega)$, it is sufficient to compute $G(u)$ for large values of $u$. In the limit $Ju\gg1$ we can use a stationary phase argument and consider only the small $p$ contribution of $|q_p(\tau)|^2$ to the integrals in Eq. \eqref{caratteristicalavoro}. The small $p$ expansion of $|q_p(\tau)|^2$ can be straightforwardly computed from Eq. \eqref{soluzione}. This computation differs from the sudden quench case \cite{Silva} only in the expression of $q_p(\tau)$; while in the latter $q_p(\tau)$ is time-independent, in this case it is a complicated expression depending on the details of the protocol. On the other hand, the squareroot singularity at $2\Delta_f$ is left unchanged. Apart from this important difference, the computation of  $P(\omega)$ follows a standard procedure, see for instance \cite{Silva}. We mention that a similar technique has been developed in \cite{Smacchia} to compute the statistics of the work done by globally changing in time the mass in a free bosonic field theory with relativistic dispersion and for generic time variations of the transverse field in a Quantum Ising Chain.

\end{appendix}


\begin{thebibliography}{30}
\bibitem{notaexp} Expansion \eqref{expansion} can be truncated at the first order provided that $\Gamma \ll 2\Big|\frac{\partial_g E_{GS}(g)}{\partial_g^2E_{GS}(g)}\Big|\equiv \mathcal{F}(g)$. $\mathcal{F}(g)$ is an increasing positive monotonic function that never vanishes in the paramagnetic phase ($g>1$).
\bibitem{nota2} The function $\varXi(g)$ is a monotomic function within the paramagnetic phase ($g>1$), $0.45\lesssim \varXi(g)\lesssim0.50$.
\bibitem{nota7} This can be easily seen introducing a finite correlation time $\tau_c$, $\langle\eta(t)\eta(t')\rangle=\frac{\Gamma}{2}\frac{1}{\sqrt{2\pi\tau_c^2}}e^{-\frac{(t-t')^2}{2\tau_c^2}}$; taking the limit $t\rightarrow t'$, from the two-time correlation function we need to recover $\langle \eta^2(t)\rangle\sim\gamma^2$, and so $\gamma\sim\sqrt{\frac{\Gamma}{\tau_c}}$. If the condition $\tau\gg\tau_c$ is properly taken into account, a gaussian correlated noise gives the same result of a delta correlated noise.
\bibitem{nota3} The saddle-point method efficiently approximates the value of the integral in the $\Gamma t\gg1$ limit, if $\Gamma t\gg(g+1)^2$.
\bibitem{note1}
For $\Gamma/\Delta \ll 1$ vertex corrections entering the self energies as well as the correlation functions can be neglected.
\bibitem{nota5} In order to make $\rho_{xx}(R)$ identically equal to the Toeplitz determinant, $T_n$, it is necessarry to shift $\rho_{xx}(R)$ from $R$ to $R+1$ and this causes the appearance of a factor $e^{-ik}$ in the expression of $s(R)$.
\bibitem{nota6} Changing the integration circle from unitary radius to radius $1/\lambda$,  we get $c_p=\int_{-\pi}^{\pi}\frac{d\theta}{2\pi}f(\theta)e^{-ip\theta}=\lambda^p\int_{-\pi}^{\pi}f(\frac{e^{i\theta}}{\lambda})e^{-ip\theta}\frac{d\theta}{2\pi}$. For the asymptotics of the Toeplitz determinant, we are interested in  $ l_p=\int_{-\pi}^{\pi}\log f(\theta)e^{-ip\theta}\frac{d\theta}{2\pi}=\oint_{|z|=1}\log f(z)z^{-p}\frac{dz}{2\pi iz} =\oint_{|z|=1/\lambda}\log f(z)z^{-p}\frac{dz}{2\pi iz}\substack{\ \\[1.5mm]=\\ z'=z\lambda}\oint_{|z'|=1}\log f\Big(\frac{z'}{\lambda}\Big)\frac{z'^{-p}}{\lambda^{-p}}\frac{dz'}{2\pi iz'}=\int_{-\pi}^{\pi}\log\Big(\frac{e^{i\theta}}{\lambda}\Big)\frac{e^{-ip\theta}}{\lambda^{-p}}\frac{d\theta}{2\pi}$, where in the last passage we used $z'=e^{i\theta}$ and so $l_0=\int_{-\pi}^{\pi}\log\frac{e^{i\theta}}{\lambda}\frac{d\theta}{2\pi}$ (see also \cite{Frankel}).
\end{thebibliography}
\end{document}